\documentclass[prd,aps,
                twocolumn,
                a4paper,floatfix,nofootinbib]{revtex4-2}

\usepackage{graphicx}% Include figure files
\usepackage{dcolumn}% Align table columns on decimal point
\usepackage{bm}% bold math
\usepackage{amsmath}
\usepackage[usenames,dvipsnames,table,xcdraw]{xcolor}
\usepackage{xspace}
\usepackage{array}
\usepackage{booktabs}
\usepackage{xcolor}
\usepackage{amssymb}% http://ctan.org/pkg/amssymb
\usepackage{pifont}% http://ctan.org/pkg/pifont
\usepackage{multirow}
\usepackage{diagbox}
\usepackage{comment}
\usepackage{xcolor}
\usepackage{float}
\usepackage{enumerate}
\usepackage{makecell}
\usepackage{booktabs}
\usepackage[normalem]{ulem}
\usepackage[utf8]{inputenc}
\usepackage{hyperref}
\allowdisplaybreaks[1]
\usepackage{units}
\usepackage{enumerate}
\usepackage{enumitem}
\hypersetup{
    colorlinks = true,
    linkcolor = {blue},
    citecolor = {blue},
    urlcolor = {blue},
    linkbordercolor = {white},
    }
\graphicspath{ {./Figures/} }
\def\d{\mathrm{d}}

\DeclareMathOperator*{\argmin}{argmin} % no space, limits underneath in displays

\def\ttrans{t_{\mathrm{trans}}}
\def\ftrans{f_{\mathrm{trans}}}
\def\ellm{(\ell m)}

\newcommand{\runtime}[3]{%
    \makebox[1.6em][r]{#1d}%
    \makebox[2.2em][r]{#2h}%
    \makebox[3.3em][r]{#3min}%
}

\newcommand{\tmerge}{{t_\mathrm{merger}}}
\newcommand{\keff}{{k_l^\mathrm{eff}}}
\newcommand{\dt}{\Delta t}

\begin{document}

\def\FD{{\texttt{FD}\ }}
\def\vfivet{{\texttt{SEOBNRv5T}}}
\def\vfivethm{{\texttt{SEOBNRv5THM}}}
\def\vfivetfd{{\texttt{SEOBNRv5T\_FD}}}
\def\vfivethmfd{{\texttt{SEOBNRv5THM\_FD}}}
\def\vfivetp{{\texttt{SEOBNRv5TPHM}}}
\def\vfourt{{\texttt{SEOBNRv4T}}}
\def\vfivehm{{\texttt{SEOBNRv5HM}}}
\def\vfiveHM{{\texttt{SEOBNRv5HM}}}
\def\vfive{{\texttt{SEOBNRv5}}}

\def\nrtidalvthree{{\texttt{NRTidalv3}}}
\def\nrtidal{{\texttt{NRTidal}}}
\def\nrtidalvtwo{{\texttt{NRTidalv2}}}
\def\nrtvthree{{\texttt{NRTidalv3}}}
\def\imrphenomdnrtidaltwo{{\texttt{IMRPhenomD\_NRTidalv2}}}
\def\imrphenomdnrtidalthree{{\texttt{IMRPhenomD\_NRTidalv3}}}
\def\imrphenomxasnrtidaltwo{{\texttt{IMRPhenomXAS\_NRTidalv2}}}
\def\xasnrtidalvthree{{\texttt{IMRPhenomXAS\_NRTidalv3}}}
\def\imrphenomxpnrtidalthree{{\texttt{IMRPhenomXP\_NRTidalv3}}}
\def\imrphenomxpnrtidaltwo{{\texttt{IMRPhenomXP\_NRTidalv2}}}
\def\imrphenompvtwonrtidaltwo{{\texttt{IMRPhenomPv2\_NRTidalv2}}}
\def\vfourT{{\texttt{SEOBNRv4T}}}
\def\vfiveT{{\texttt{SEOBNRv5THM}}}
\def\lalsuite{{\texttt{LALSuite}}}
\def\seobnrvfiverom{{\texttt{SEOBNRv5\_ROM}}}
\def\seobnrvfourromnrtidalvtwo{{\texttt{SEOBNRv4\_ROM\_NRTidalv2}}}
\def\seobnrvfiveromnrtidalvtwo{{\texttt{SEOBNRv5\_ROM\_NRTidalv2}}}
\def\seobnrvfiveromnrtidalvthree{{\texttt{SEOBNRv5\_ROM\_NRTidalv3}}}
\def\teob{{\texttt{TEOBResumS}}}
\def\teobg{{\texttt{TEOBResumS-Giotto}}}
\def\teobspa{{\texttt{TEOBResumSPA}}}
\def\SPA{{\mathrm{SPA}}}

\title{Speed and accuracy for long signals: Frequency-domain effective-one-body waveforms for compact binary coalescences}

\author{Marcus \surname{Haberland}$^{1}$}\email{marcus.haberland@aei.mpg.de}
\author{Alessandra \surname{Buonanno}$^{1,2}$}\email{alessandra.buonanno@aei.mpg.de}

\affiliation{${}^1$Max Planck Institute for Gravitational Physics (Albert Einstein Institute), Am M\"uhlenberg 1, Potsdam 14476, Germany}
\affiliation{${}^2$Department of Physics, University of Maryland, College Park, Maryland 20742, USA}

\date{\today}

\begin{abstract}
Gravitational-wave inference for long signals, like those from binary neutron-star (BNS) systems, requires waveform models that are both physically faithful and computationally efficient, otherwise, one risks drawing incorrect conclusions about nuclear matter from observations.
To address this challenge, we present a frequency-domain implementation of the accurate \texttt{SEOBNRv5THM} waveform model for quasi-circular, spin-aligned BNS systems within the effective-one-body framework. Our approach combines the stationary-phase approximation (SPA) for the early inspiral with a fast Fourier transform treatment of the late- and post-inspiral regime, applied mode-by-mode. Our hybrid approach retains the efficiency of the SPA without affecting the waveform accuracy close to merger, where matter effects are most significant.
The resulting waveform's generation speed can be further decreased using modern parameter-estimation techniques, such as multibanding and relative binning. We demonstrate excellent agreement with the baseline \texttt{SEOBNRv5THM} model in both mismatches and when analyzing real and synthetic data, and show how waveform systematics could affect BNS detections in upcoming observational runs and new facilities on the ground. We find that our method significantly reduces computational costs, enabling faithful parameter estimation for BNS signals within practical runtimes of $\mathcal{O}$(days). Our procedure can be readily extended to coalescing binary black hole systems.
\end{abstract}

\maketitle

\section{Introduction} \label{section:Introduction}

The field of gravitational-wave (GW) astronomy has undergone rapid expansion since the first detection of a binary--black-hole (BBH) merger in 2015~\cite{LIGOScientific:2016aoc}, and the subsequent observation of a binary--neutron-star (BNS) merger in 2017~\cite{LIGOScientific:2017vwq,LIGOScientific:2018hze,Dietrich:2020efo}. The latter, GW170817, marked the beginning of multi-messenger GW astronomy, enabling joint gravitational and electromagnetic observations and providing unprecedented insights into astrophysics, fundamental physics, and cosmology~\cite{LIGOScientific:2017vwq, LIGOScientific:2017zic, LIGOScientific:2017ync}. Since then, the LIGO--Virgo--KAGRA (LVK) Collaboration has reported over 200 candidates for compact-binary detections~\cite{LIGOScientific:2025slb,GWTC3,GWTC2.1}, including another BNS signal in GW190425~\cite{LIGOScientific:2020aai}, with further observations from the remainder of the fourth observing run (O4) currently being analyzed.

The analysis of the first part of data from this run (denoted as O4a) has brought us no new significant BNS detections~\cite{LIGOScientific:2025slb} apart from a sub-threshold trigger~\cite{Niu:2025nha}, leaving the two aforementioned BNS events as the only confident BNS mergers observed so far.
This continues the reduction of the most likely BNS merger rate in the local universe, which is currently being assumed to be $\mathcal{O}(100)\ \mathrm{Gpc}^{-3}\mathrm{yr}^{-1}$~\cite{LIGOScientific:2021psn,LIGOScientific:2025pvj,Fishbach:2026vxv,Salafia:2025qva}, with large uncertainties remaining, implying that it could be larger or smaller by a factor of a few.
Even at their current sensitivities, BNS detections in LVK detectors are still governed by small-number Poissonian statistics, leaving significant probability for non-detections over individual observing runs. The absence of BNS mergers in O4a is therefore consistent with previous expectations~\cite{LIGOScientific:2021psn}, although there is a growing tension with astrophysical predictions for the BNS merger rates due to the continued non-observation~\cite{Fishbach:2026vxv}. Even when accounting for these non-detections, the BNS range in the upcoming LVK six-month observing run, designated Intermediate Run 1 (IR1), is still expected to be around $170\ \mathrm{Mpc}$ for the individual LIGO detectors~\cite{ObservingScenarioTimeline}. Assuming a BNS merger rate of $100\ \mathrm{Gpc}^{-3}\mathrm{yr}^{-1}$, this corresponds to an expected number of BNS detections of $N_{\mathrm{det}}\simeq 4\pi/(3)\,(0.17\,\mathrm{Gpc})^3\,\left(100\,\mathrm{Gpc}^{-3}\,\mathrm{yr}^{-1}\right)\,(0.5\,\mathrm{yr})\simeq 1\,,$
with a corresponding 64\% chance of detection under a Poisson distribution\footnote{For a more realistic detector network, one can also consider the duty cycles of the individual detectors and that the BNS ranges of the individual detectors approximately add in quadrature, which changes the number of expected detections slightly upwards by a factor $\sim2$, see also Refs.~\cite{Salafia:2025qva,Kiendrebeogo:2023hzf,Weizmann:2026xyz}.}. As the expected BNS range of individual detectors in the fifth observing run (O5)~\cite{ObservingScenarioTimeline,LVK_psds} is instead on the order of 300~Mpc, this results in about ten expected BNS detections per year.

In the more distant future, this situation is anticipated to change with the deployment of next-generation detectors, such as the Einstein Telescope (ET) and Cosmic Explorer (CE). Given their much higher sensitivities, and the potential employment of both of these at the same time, one expects an increased BNS detection rate of around ten-thousand events per year, with signal-to-noise ratios (SNR) above one hundred for around a hundred of them per year~\cite{Iacovelli:2022,Gupta:2023lga,Kalogera:2021bya,Reitze:2019iox,Evans:2021gyd,ET:2025xjr}\footnote{The non-detections of BNS in O4a are not in tension with these expectations, as the quoted results for ET and CE are based on a realistic BNS merger rate of $100\ \mathrm{Gpc}^{-3}\mathrm{yr}^{-1}$ (see, e.g., Refs.~\cite{Iacovelli:2022,Gupta:2023lga}). Similarly to the previous discussion, this might however decrease by a factor of a few if the non-detections continue through O5.}.

In both the near and far future, significant insights into the interiors of NSs, as well as cosmology, may therefore be obtainable with upcoming detections of the GW emission of BNS, going beyond the conclusions drawn from GW170817. For one, the matter in the core of a NS is a few times denser than the nuclear saturation density~\cite{Lattimer:2012nd, Ozel:2016oaf}, and it cannot be constrained in earth-based experiments due to its extreme pressure and internal energy. Consequently, there are various realistic theoretical and phenomenological models in nuclear physics (see, e.g., Refs.~\cite{Chin:1974sa, Serot:1997xg, Huth:2021bsp, Alford:2022bpp, Lattimer:2012nd, Lattimer:2021emm, Burgio:2021vgk, Zhu:2023ijx}), which predict the state of such matter and would constrain the high-energy limit of nuclear physics if confirmed or rejected.
Additionally, BNS coalescences can be used as standard sirens to measure the Hubble constant $H_0$ and other cosmological constants as bright or Love sirens~\cite{Schutz:1986gp, LIGOScientific:2017adf,Dhani:2022ulg}, respectively.

These applications rest on an accurate modeling of the individual NSs during their inspiral and merger~\cite{Flanagan:2007ix, Hinderer:2009ca, LIGOScientific:2017vwq, LIGOScientific:2018hze, LIGOScientific:2020aai}. The interior structure of NSs, dictated by its equation of state (EoS), determines how it reacts to a time-varying tidal field of a compact binary partner. Consequently, one can define the NS's EoS-dependent tidal deformability $\Lambda_2$, which quantifies the star's quadrupolar deformation in response to an external tidal field~\cite{Flanagan:2007ix, Hinderer:2007mb}. An important goal of BNS detection is then to determine this deformability as well as the individual star's mass, spin, and potentially other NS quantities, to determine the correct EoS. Elaborating on one of these NS parameters, it has been predicted that during the late inspiral, the frequency of the time-varying tidal field can approach or cross the fundamental $f$-mode frequency of a NS. This can result in resonance phenomena that further augment the star's tidal response. These frequency-dependent dynamical tides have been studied in Refs.~\cite{1994PThPh..91..871S, Flanagan:2007ix, Hinderer:2016eia, Steinhoff:2016rfi,Schmidt:2019wrl, Andersson:2019dwg, Gupta:2020lnv, Steinhoff:2021dsn, Pratten:2021pro, Kuan:2022etu, Pnigouras:2022zpx, Mandal:2023lgy, Mandal:2023hqa} and can be effectively described through separation-dependent tidal enhancement $\Lambda_l(r)=\keff(r)\Lambda_l$ and depend through the resonance frequency $\omega_{f,l}$ on the EoS as well as the NS's spin. In addition to these changes of the dynamical tides, the spin also induces a multipolar structure in BHs and NSs, which can be captured by higher-order spin effects~\cite{Henry:2022dzx,Khalilv5}.

Given the higher SNRs, there is furthermore growing evidence that waveform systematics might be significant for BNS systems with upcoming observing runs and future detectors~\cite{Samajdar:2018,Gamba:2021prd,Kunert:2022,Kunert:2024,Huez:2025npe}. To reliably draw conclusions from future BNS detections, significantly more accurate waveform models will be required.
A historically very successful waveform-modeling-approach that combines results from different analytical approximation methods with information from high-accuracy NR simulations and strong-gravity effects is the effective-one-body (EOB) formalism (see, e.g., Refs.~\cite{Buonanno:1998gg,Buonanno:2000ef,Buonanno:2006ui,Damour:2000we,Damour:2001tu,Buonanno:2005xu}). It maps the dynamics of a compact binary to that of a test mass in a deformed Kerr 
background, with the deformation parameter being the symmetric mass ratio $\nu$, supplemented with an associated energy loss due to GW emission. The resulting dissipative system of ordinary differential equations is then integrated numerically. As a last step, the resulting waveform is computed from the trajectories.
EOB models are developed in two flavors, namely \texttt{SEOBNR}\ in Refs.~\cite{Bohe:2016gbl,Cotesta:2018fcv,Ossokine:2020kjp,Cotesta:2020qhw,Ramos-Buades:2021adz,Mihaylov:2021bpf,Hinderer:2016eia,Steinhoff:2016rfi,Haberland:2025luz}
and \teob\ in Refs.~\cite{Nagar:2018zoe,Bernuzzi:2014owa,Nagar:2019wds,Nagar:2020pcj,Gamba:2021ydi,Riemenschneider:2021ppj,Chiaramello:2020ehz,Bernuzzi:2014owa,Akcay:2018yyh,Gonzalez:2022prs,Gamba:2023mww} and both flavors have also been extended to BNS systems~\cite{Hinderer:2016eia,Steinhoff:2016rfi,Haberland:2025luz,Bernuzzi:2014owa,Nagar:2018zoe,Akcay:2018yyh,Gonzalez:2022prs,Gamba:2023mww}. \teob\ is furthermore developed as \teobg\ for quasi-circular systems, and as \texttt{TEOBResumS-Dal\'i} for eccentric systems.

In this work, we focus on \vfivethm, the quasi-circular, aligned-spin BNS waveform model of Ref.~\cite{Haberland:2025luz}, which represents a significant advancement over the previous model \vfourt~\cite{Steinhoff:2016rfi,Lackey:2018zvw} in both speed and accuracy. Its key improvements include
\begin{enumerate}[label=(\roman*)]
    \itemsep-3pt
    \item the state-of-the-art BBH model \vfivehm~\cite{Pompiliv5} as the point-particle baseline, which uses the Hamiltonian of a test-mass (instead of a test-spin) in a deformed Kerr spacetime,
    \item spin-shifted dynamical tides, which are included through an effective enhancement factor $\keff(r;\chi)$ of the adiabatic tidal deformability $\Lambda_l$~\cite{Steinhoff:2016rfi}, in particular including the shift of the $f$-mode frequency and resonance response due to low spins~\cite{Steinhoff:2021dsn}, that is not included in other state-of-the-art BNS waveform models,
    \item the consistent inclusion of adiabatic tidal information up to 7.5PN order~\cite{Henry:2020ski,Mandal:2024,Dones:2024} and spin-induced multipole moments up to 3.5PN order~\cite{Henry:2022dzx,Khalilv5},
    \item the development of a pre-merger model and calibration to NR waveforms across a wide range of tidal deformabilities with mass ratios up to $q = 2$,
    \item the inclusion of higher modes, namely the dominant (2,2) mode and the largest subdominant modes: (3,3), (2,1), (4,4), (3,2), (5,5), and (4,3), and
    \item the usage of the post-adiabatic approximation~\cite{Nagar:2018gnk,Mihaylov:2021bpf} among other computational improvements, which already reduces the time-domain waveform generation time by 100 to 1000 times.
\end{enumerate}

The gravitational radiation of a binary coalescence can be decomposed into spin-weighted spherical harmonic modes $h_{\ell m}(t)$, with the dominant mode being the (2,2)-mode. For BNS systems, with their similar masses and spins, the higher-order modes are expected to be suppressed. Given that they are a physical effect, they should however not be neglected in all cases, as they can be important for accurate parameter estimation (PE), in particular for the binary's mass ratio and inclination angle~\cite{Cotesta:2018fcv,Varma:2019csw}.
For BNS coalescences, matter effects furthermore affect the dynamics and modify the waveform close to merger, in particular in the last few orbits~\cite{Dietrich:2017}. This leads to inconsistencies with respect to NR if neglected. In \vfivethm, the pre-merger part of the dominant (2,2)-mode is therefore phenomenologically modeled to agree with NR to a high amount of accuracy.

An increase in accuracy of the predicted GW signal should however not come at the cost of waveform generation speed, as PE of BNS signals poses a significant computational challenge. Standard Bayesian inference methods for PE~\cite{Veitch:2014wba,Thrane:2018qnx} rely on repeated comparison of the observed data with simulated signals in the frequency domain. There, the detector noise is well approximated as stationary and Gaussian, and one can easily marginalize over detector calibration uncertainties that are typically provided as frequency domain cubic-spline priors. Coalescing BNS systems, with their low total masses $M \sim 3\ M_{\odot}$, remain in the detector band for thousands of wave-cycles, which correspond to multiple minutes in LVK detectors. Given that the detectors typically sample data at 16~kHz, the number of data points per signal is therefore on the order of $\mathcal{O}(10^6)$ and correspondingly orders of magnitude larger than for typical BBH signals. Going forward, this situation will become even worse with the multiple-hour long signals expected in ET and CE. In this case, individual BNS signals correspond to $\mathcal{O}(10^8)$ data points (and therefore gigabytes of data per event and waveform evaluation).

To mitigate this challenge, several strategies have been developed. Most notable are the multibanding~\cite{Vinciguerra_2017,Morisaki:2021ngj,Cornish:2021lje} and relative binning~\cite{Zackay:2018qdy,Krishna:2023bug,Leslie:2021ssu} techniques, which reduce the number of frequency evaluations required for likelihood computations by orders of magnitude without affecting the posteriors. 
Surrogate models~\cite{Hu:2024lrj,Iwaya:2026feu} and PE codes based on machine-learning methods, notably neural posterior estimation~\cite{Dax:2021tsq,Tissino:2022thn,Dax:2022pxd,Dax:2024}, are also under continuous development to speed up inference studies and utilize the previously mentioned techniques. 
These approaches however require frequency-domain waveform models that can be evaluated on sparse, non-uniform frequency-grids. The most used BNS modeling scheme in this class is the phenomenological \nrtidal\ family of waveform models, developed in Refs.~\cite{Dietrich:2017aum, Dietrich:2019kaq,Kawaguchi:2018gvj,Abac:NRT}, with its current version being \nrtidalvthree~\cite{Abac:NRT,Abac:2025brd}. It aims to model the tidal phase contribution during the inspiral through the use of closed-form analytical expressions for the tidal contribution calibrated to NR. The matter contribution is then used to augment a given BBH waveform baseline to model BNS systems.

Time-domain waveform models like EOB, on the other hand, generally rest on fewer modeling assumptions and continue a proven track record of accuracy. Their main drawback is that they usually need to be interpolated to an equidistant time-grid and transformed to the frequency domain via a subsequent FFT, which becomes computationally expensive for long signals and prohibits the aforementioned PE techniques.
As an alternative, the stationary-phase approximation (SPA) has long been used to map inspiral waveforms to the frequency domain efficiently~\cite{Sathyaprakash:1991mt,Droz:1999qx,Gamba:2020ljo}. In the context of EOB, this has been done most notably in the \teobspa\ extension to \teobg\ in Ref.~\cite{Gamba:2020ljo}, which can also be used with relative binning and multibanding~\cite{Huez:2025gja}. The validity of the SPA is however limited to regimes where the signal evolves adiabatically. It therefore becomes inaccurate near merger, where the frequency evolution becomes rapid and potentially non-monotonic, which can be even more true for the higher-order modes. This is particularly detrimental, as the last few orbits are where matter effects become most pronounced, and where we would like to measure them most accurately. There also exists the extension of the SPA to the shifted uniform asymptotics, which mitigates some, but not all, of these issues~\cite{Klein:2013qda,Klein:2014bua,Morras:2025nlp}.
Surrogate modeling or reduced-order-methods~\cite{Lackey:2018zvw,Cotesta:2020qhw} can also be used to speed up time-domain models, but they lack the flexibility to incorporate new physical information as it is uncovered analytically or via numerical simulations. They additionally involve significant upfront costs to build the surrogate model.

In this paper, we develop a frequency-domain version of \vfivethm\, following previous work~\cite{Cornish:2020odn,Gamba:2020ljo}. Our approach combines the strengths of analytical and numerical methods by applying the SPA to the early inspiral, where it is valid and efficient, and complementing it with an FFT-based treatment of the late inspiral and pre-merger signal. Crucially, this construction is performed mode-by-mode and enables the direct evaluation of the waveform on arbitrary frequency grids, making it naturally compatible with modern PE techniques such as multibanding and relative binning.

The remainder of this paper is organized as follows. In Sec.~\ref{sec:FrequencyDomain}, we describe the hybrid SPA--FFT approach in detail. In Sec.~\ref{sec:mb_rb}, we review and apply modern PE techniques for long signals. In Sec.~\ref{sec:speed_robustness}, we present performance and accuracy results, in order to assess the applicability of our model in the context of PE, which we perform in Sec.~\ref{sec:pe}. We conclude in Sec.~\ref{sec:conclusion} with a summary and outlook.

\begin{figure*}[t]
    \centering
   \includegraphics[width=0.95\linewidth]{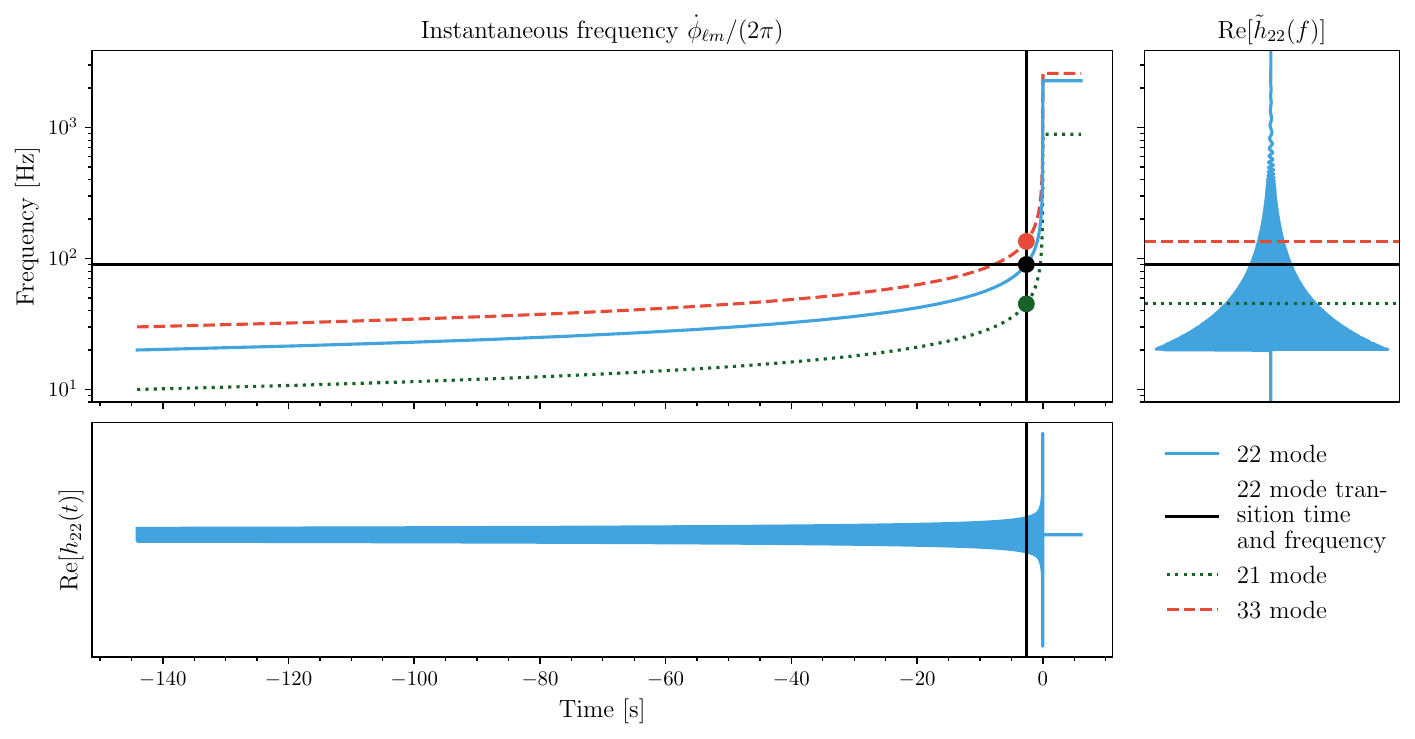}
   \caption{A pedagogical comparison of how we generate the frequency-domain waveform via the SPA and FFT on a per-mode basis. We show the waveform in time domain (bottom), the instantaneous frequency evolution of the individual modes (middle) and the frequency-domain waveform (right). The SPA is applied for early times and low frequencies, while the FFT is applied for late times and high frequencies. We indicate how the same transition time corresponds to different transition frequencies for the different modes. Note that the transition time is chosen for illustration purposes and is actually much closer to the merger time in the actual model. It is also allowed to vary between the individual modes.}
   \label{fig:pedagogical}
\end{figure*}

\section{Building a flexible frequency-domain model} \label{sec:FrequencyDomain}

We now describe how we build a frequency-domain version of the \vfivethm\ model, which we refer to as \vfivethmfd. The key idea is to combine the strengths of analytical and numerical methods by applying the SPA to the early inspiral, where it is valid and efficient, and complementing it with an FFT-based treatment of the late inspiral and pre-merger signal.
We perform the transition from the SPA to the FFT on a per-mode basis, by choosing a transition-time $\ttrans^{\ellm}$ that corresponds to a transition-frequency $\ftrans^{\ellm}$ until which the SPA is valid. After that time and above this frequency, we interpolate the modes to a uniform time-grid to perform an FFT to faithfully capture the post-inspiral part of the mode.
As a final step, we then combine the resulting frequency-domain waveforms on a flexible frequency grid and evaluate the frequency-domain polarizations on it. In Fig.~\ref{fig:pedagogical}, we show for an exemplary waveform how one can transition from the time domain to the frequency domain via the SPA and FFT on a per-mode basis, and how the same transition time corresponds to different transition frequencies for the different modes.

In Sec.~\ref{subsec:SPA}, we elaborate on all of the above steps in detail and explain how we develop a frequency-domain version of the \vfivethm\ model for flexible frequency grids, and describe how the inspiral waveform modes are computed in the frequency domain via the SPA. In Sec.~\ref{subsec:FFT}, we then explain how we compute the late-inspiral-pre-merger waveform modes in the frequency domain via an FFT. We finally conclude on how we combine the two methods to arrive at the frequency-domain strain $\tilde{h}_{+,\times}(t)$ in Sec.~\ref{subsec:combination}. In the following, we use natural units in which $c = G = 1$ unless stated otherwise, and we use overhead dots to signal time derivatives.

\subsection{The SPA for the early inspiral} \label{subsec:SPA}

The complex linear combination of GW polarizations, $h(t) \equiv h_{+}(t) -ih_{\times}(t)$, are determined from the EOB dynamics and can be expanded in the basis 
of $-2$ spin-weighted spherical harmonics as follows~\cite{Pan:2011gk,Pompiliv5,Haberland:2025luz}:
\begin{equation}
	h(t;  \bm{\lambda}, \iota, \varphi) = \sum_{\ell \geq 2}\sum_{|m|\leq \ell} {}_{-2} Y_{\ell m} (\iota, \varphi) \, h_{\ell m}(t;\bm{\lambda}),
\label{eq:hoft_sphericalH}
\end{equation}
where $\bm{\lambda}$ denotes the intrinsic parameters of the compact binary, such as mass ratio $q$, spins $\chi_{1,2}$, and the NS parameters $(\Lambda_l, \omega_l, \Delta \omega_l, C_l)$. They correspond to the tidal deformability, the $f$-mode frequency, the shift of it due to spin, and the spin-induced multipole moments, respectively.
The parameters $(\iota, \varphi)$ describe the binary's inclination angle (computed with respect to the 
direction perpendicular to the orbital plane) and the azimuthal direction to the observer, respectively.

We define the signal to occur over the interval $t\in[0,T]$, with $T$ being the signal duration in our detector. We remind again that the polarizations are real-valued functions, while the modes $h_{\ell m}$ are complex-valued functions, which behave as $h_{\ell m}(t) = a_{\ell m}(t)e^{-i \phi_{\ell m}(t)}$.
Due to the azimuthal symmetry for aligned-spin binaries $h_{\ell m}=(-1)^{\ell} h_{\ell-m}^*$, the phases of the modes behave differently based on their magnetic number $m$. In particular, $m>0$ modes have a monotonically increasing phase, and the $m<0$ modes have a monotonically decreasing phase. For now, we focus on the $m>0$ modes with monotonically increasing frequency $\omega_{\ell m} \equiv \dot{\phi}_{\ell m}>0$ as is the convention of the \texttt{SEOBNR} family of aligned-spin waveform models. The $m<0$ modes are then computed via the azimuthal symmetry.

To compute our modes in the frequency domain, we employ the standard convention of the Fourier transform
\begin{equation} \label{eq:Fourier_transform}
    \tilde{h}_{\ell m}(f) = \int_{-\infty}^\infty \d t \, h_{\ell m}(t)\, e^{-2i \pi f t}\ ,
\end{equation}
and follow the derivation of the SPA as outlined in Refs.~\cite{Sathyaprakash:1991mt,Droz:1999qx,Gamba:2020ljo,Cotesta:2020qhw}. Introducing the previous assumptions on our modes, we can rewrite our signal as
\begin{subequations}
    \label{eq:htilde}
    \begin{align}
	\tilde{h}_{\ell m}(f) &= \int_{0}^{T} \d t\, a_{\ell m}(t) e^{-i[\phi_{\ell m}(t)+2 \pi f t]} \\
    &= \int_{0}^{T} \d t\, a_{\ell m}(t)\, e^{-i \psi_f^{\ellm}(t)}\,,
    \end{align}
\end{subequations}
with
\begin{equation}
\psi^{\ellm}_f(t) \equiv \phi_{\ell m}(t) + 2 \pi f t \ .
\end{equation}

Since the integrand oscillates rapidly, the largest contribution to
the integral comes from the vicinity of the stationary points of the 
phase $\psi^{\ellm}_f(t)$ (i.e., when $\dot{\psi}^{\ellm}_f(t)=0$). Defining these saddle points of $ \psi_{f}^{\ellm}(t)$ as $t_f$, we find their location to be at 
\begin{equation}
    \label{eq:saddle_point}
    \omega_{\ell m}(t_f) = - 2\pi f > 0\ ,
\end{equation}
which implicitly defines a time-frequency correspondence $t_f = t(f)$. 
Given our sign convention and the monotonicity of the phase, stationary points in the $m>0$ ($m<0$) waveform modes only exist for negative frequencies $f<0$ (positive frequencies $f>0$), and we can therefore neglect the positive frequencies (negative frequencies) in the Fourier transformed signal, which will be approximately zero.

By Taylor expanding the integrand in Eq.~\eqref{eq:htilde} as
\begin{subequations} \label{eq:psi_approx}
\begin{align}
  \psi_{f}^{\ellm}(t) & \simeq \psi^{\ellm}_{f}\left(t_{f}\right)+
  \frac{\dot{\omega}_{\ell m}(t_f)}{2}\left(t-t_{f}\right)^{2} \ ,\\
  a_{\ell m}(t) & \simeq a_{\ell m}\left(t_{f}\right) \ ,
\end{align}
\end{subequations}
one finds the Gaussian integral
\begin{equation}
    \begin{aligned}
        \tilde{h}_{\ell m}(f) \simeq &\ a_{\ell m}(t_f) e^{-i[\phi_{\ell m}(t_f) + 2 \pi f t_f]} \\ 
        & \times \int_{-\infty}^{\infty} \d t\, e^{-\frac{i}{2}\dot{\omega}_{\ell m}(t_f)\left(t-t_{f}\right)^{2}}\ .
    \end{aligned}
    \label{eq:gaussian}
\end{equation}
This can easily be solved to obtain 
\begin{equation} \label{eq:SPA}
\tilde{h}^{\rm SPA}_{\ell m}(f) = A_{\ell m}^{\rm SPA}(f) e^{-i \Psi_{\ell m}^{\rm SPA}(f)}
\end{equation}
with
\begin{subequations} \label{eq:SPA_components}
\begin{align}
	A_{\ell m}^{\rm SPA}(f) &= a_{\ell m}(t_f) \sqrt{\frac{2 \pi}{\dot{\omega}_{\ell m}(t_f)}} \,, \label{eq:SPA_amplitude} \\
	\Psi_{\ell m}^{\rm SPA}(f) &= \phi_{\ell m}(t_f) + 2\pi f t_f + \frac{\pi}{4} \\ 
    &= \phi_{\ell m}(t_f) - \omega_{\ell m}(t_f) t_f + \frac{\pi}{4}
    \,, \nonumber
\end{align}
\end{subequations}
where the last line follows from Eq.~\eqref{eq:saddle_point}, which is more convenient to work with, as everything is expressed in terms of $t_f$ and can therefore easily be evaluated directly from the time-domain waveform. The resulting Fourier transformed signal at a given frequency $f<0$ is then fully determined by points in the waveform where the instantaneous GW frequency $\omega_{\ell m}/(2\pi)$ sweeps through $|f|$ in the time domain.

\subsubsection{Determining the end of the SPA waveform} 

The above derivation of the SPA relies on multiple implicit assumptions which we can use to inform when to apply the transition to the FFT method (i.e., when to end the SPA part of the waveform model). For Eqs.~\eqref{eq:psi_approx} to be accurate, one for example needs to assume that the first order expansion of $\psi_f^{\ellm}$ and the constant amplitude term are sufficient over the time-scale of the integration. In the literature, one therefore usually sees a comparison of the next orders in the Taylor expansion to the timescales of change, which leads to the conditions (see, e.g., Ref.~\cite{Cotesta:2020qhw}\footnote{In this reference, also a third condition $\left| \dot{a}_{\ell m} / (a_{\ell m}\omega_{\ell m})\right| \ll 1$ is stated, which is however redundant and implied by the other conditions we state, as they together imply $\left| \dot{a}_{\ell m} \right| \ll \left| a_{\ell m}\sqrt{\dot{\omega}_{\ell m}} \right| \ll \left| a_{\ell m} \omega_{\ell m} \right|$, which is the aforementioned third condition.})

\begin{equation} \label{eq:SPA_conditions}
        \left| \frac{\dot{\omega}_{\ell m}}{\omega_{\ell m}^{2}}\right| \ll 1,\ \mathrm{ and }\ 
        \left| \frac{(\dot{a}_{\ell m} / a_{\ell m})^{2}}{\dot{\omega}_{\ell m}}\right| \ll 1.
\end{equation}
The amplitude condition on the right also implicitly includes the need of a non-vanishing $\dot{\omega}_{\ell m}$ throughout the waveform in question, which would also introduce singularities in $A_{\ell m}^{\rm SPA}$ of Eq.~\eqref{eq:SPA_amplitude}.

Another implicit condition is that the boundaries of our integration $\{0,\ T\}$ are sufficiently far from the stationary points, such that one can replace the integration boundaries by infinity, as we did going from Eq.~\eqref{eq:htilde} to Eq.~\eqref{eq:gaussian}. If this condition is not met, one needs to introduce extra Fresnel-integral terms in Eqs.~\eqref{eq:SPA_components}, which we want to avoid in our model. To this end, we remind the reader that the Gaussian integrand we study contains $\sim 99\%$ of its volume in a time interval of support
\begin{equation} \label{eq:support}
    \mathcal{S}_f = \left[t_f - 3/\sqrt{\dot\omega(t_f)},\ t_f + 3/\sqrt{\dot\omega(t_f)}\right] \subset \mathbb{R}\ ,
\end{equation}
which corresponds to three standard deviations from the mean of our Gaussian. As long as  $0<\mathcal{S}_f<\tmerge$ for a given frequency $f$, one can therefore approximately replace the integration boundaries of our Gaussian integral by infinity.

As we want to employ the FFT to keep our phenomenologically modeled and NR calibrated pre-merger waveform accurate also in the frequency domain, we set a maximum time for the SPA part of the signal of $t_{\rm max}=\tmerge-500\ M$ and stop the SPA waveform at that time the latest. We find that our pre-merger model starts to significantly change the frequency evolution of the dominant (2,2)-mode only after this time across the BNS parameter space.

We therefore end the SPA waveform either at the point of the dynamics from the ODE integration right before the frequency evolution would become non-monotonic, $\dot{\omega}_{\ell m} > 0$, at the frequency at which $\mathcal{S}_f$ first touches the merger time $t_{\rm merge}$, or before the pre-merger model sufficiently changes the dominant (2,2)-mode, so at $t_{\rm merge}-500\ M$, whichever occurs first, or algebraically
\begin{equation} \label{eq:ttrans}
\ttrans^{\ellm} = \argmin_t \left\{
\begin{array}{c}
t=t_{\rm merge} - 500\,M, \\[1ex]
t + 3/\sqrt{\dot{\omega}_{\ell m}(t)} \geq t_{\rm merge}, \\[1ex]
\dot{\omega}_{\ell m}(t+\Delta t_{\mathrm{dyn}}) = 0
\end{array}
\right\} ,
\end{equation}
where $\Delta t_{\mathrm{dyn}}$ is the time step of the EOB dynamics.
With this we can determine the transition time as $t_{\rm trans}^{(\ell m)}$ and the corresponding transition frequency $f_{\rm trans}^{(\ell m)}$.

During the development of the SPA implementation, we experimented with dynamically determining the transition point between the SPA and FFT regimes through an adaptive epsilon parameter that monitors the validity of the SPA approximation of Eq.~\eqref{eq:SPA_conditions} during waveform evaluation. However, as detailed in Sec.~\ref{subsec:mismatches}, our mismatch studies demonstrate that the mismatches introduced by our fixed transition criterion in Eq.~\eqref{eq:ttrans} remain sufficiently small across the BNS parameter space and do not significantly impact PE. Consequently, we abandoned the online epsilon determination in favor of the simpler and more computationally efficient fixed transition scheme.

For future research it is however noteworthy that we have found that the second condition of Eq.~\eqref{eq:SPA_conditions} is often the more restrictive one (i.e., that the amplitude changing over the integration time is often the most significant source of error in the SPA towards merger). This is particularly the case for the subdominant modes with $m=1$, which can have a lower $\dot{\omega}_{\ell m}$ and therefore higher error in the amplitude. It might therefore be interesting to explore how one can improve the SPA by including higher-order terms in the Taylor expansion of the amplitude, which would allow us to extend the validity of the SPA to later times during the inspiral and perhaps reduce the need for an FFT of the post-inspiral part of the waveform, as is done in the shifted-uniform asymptotics method~\cite{Klein:2013qda,Klein:2014bua,Morras:2025nlp}. 

As proof of principle for our transition scheme, in Fig.~\ref{fig:comparison}, we compare the resulting frequency-domain waveform to a conventional FFT of the entire time-domain signal, and show the agreement between the two for one exemplary system with parameters $m_1=2.4\ M_\odot,$ $m_2=1.4\ M_\odot,$ $\Lambda_2^{(1)}=400,$ $\Lambda_2^{(2)}=800,$ $\chi_1=\chi_2=0.1$. We also showcase the improvement in accuracy of the frequency-domain waveform in comparison to a sole application of the SPA, which becomes inaccurate close to merger. To this end we also compare \texttt{TEOBResumSPA} to the FFT of \texttt{TEOBResumS-Giotto}. Note how the error in both amplitude and phase increases significantly towards merger when using the SPA alone, while the error in the hybrid SPA-FFT approach goes to zero at the transition frequency, with the mismatch not increasing past this point. This is in particular the case when including higher modes, as the mismatch between the SPA and FFT waveforms increases significantly when including higher modes, while the mismatch between the hybrid SPA-FFT and FFT waveforms remains unaffected by the inclusion of higher modes.

\subsubsection{Determining the start of the SPA waveform}

\begin{figure}[t]
    \centering
   \includegraphics[width=\linewidth]{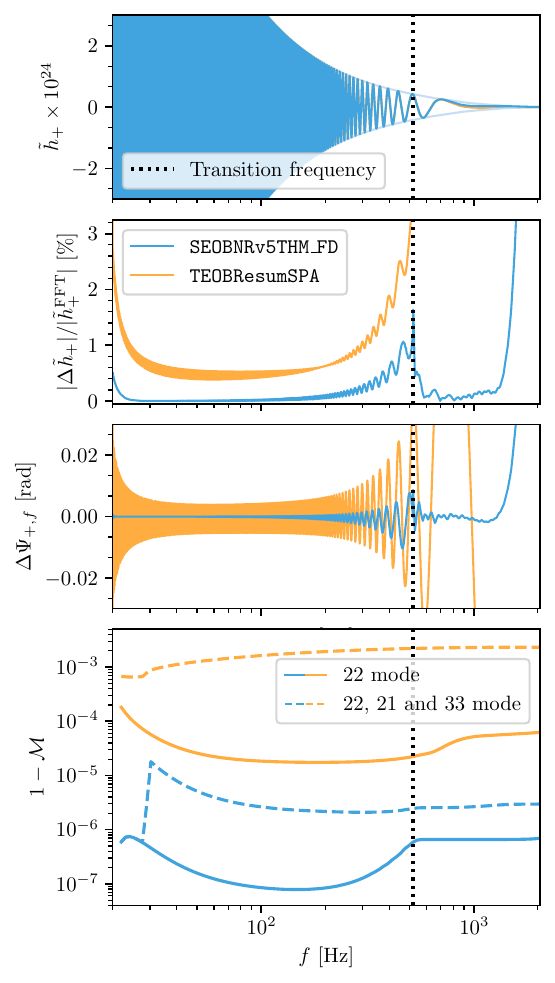}
   \caption{A showcase of the accuracy between \vfivethmfd\ and its time-domain version \vfivethm\ (blue), as well as between \teobspa\ and \teobg\ (orange) for an exemplary BNS system. Comparisons are performed with (dashed) and without (solid) higher modes as functions of frequency. We show in the different rows the frequency-domain strain with an indication of the amplitude envelope; the relative difference of the amplitudes; the absolute difference of the phase; and the optimized mismatch from Eq.~\eqref{eq:mismatch} between 20 Hz and a given maximum frequency.}
   \label{fig:comparison}
\end{figure}

Time-domain waveform models are furthermore conditioned prior to the application of an FFT, which affects the low-frequency content of the polarizations and individual waveform modes in the frequency domain. As we intend to faithfully reproduce the time-domain model in the frequency domain, we also need to carefully consider the start of the SPA waveform for different $m$ modes, which occurs at different frequencies for the individual modes (see Fig.~\ref{fig:pedagogical}).
In Fig.~\ref{fig:comparison}, one can also see the effect of this conditioning on the Fourier phase and amplitude of the different modes, through the increase in mismatch for both \teobspa\ and \vfivethmfd\ at 30 Hz, where the $m=3$ mode enters in the waveform for both approximants. Without our conditioning approach as described below, this peak is even more pronounced for our model. It should be noted however that in Ref.~\cite{Ursell:2025ufb} the bias induced by neglecting this conditioning (i.e., incorrectly assuming low-frequency power in the higher modes) was found to be stronger for high-mass and asymmetric binaries, and therefore of less importance for BNS systems. In order to ensure that our model can be used for a wider range of systems, we still want to include this conditioning in our model.

During the inspiral, the individual modes obey a simple overall scaling with the orbital phase $\phi_{\rm orb}$ as
\begin{equation}\label{eq:m_scaling}
	h_{\ell m} = \left|h_{\ell m}\right| e^{ -i\phi_{\ell m}} \simeq \left|h_{\ell m}\right| e^{ -i m \phi_{\rm orb}} \,,
\end{equation}
with this scaling deteriorating during the late-inspiral. In the notations of the EOB factorized waveforms~\cite{Damour:2008gu,Pan:2011gk}, these phase deviations come from the phases $e^{i \delta_{\ell m}}$ and tail factors $T_{\ell m}$ (see Eqs.~(14) and~(21) in Ref.~\cite{Pan:2011gk}).

Based on this, a mode with magnetic number $m$ will therefore have a frequency $f_{\rm start}^{(\ell,m)}\simeq mf_{\rm start}^{(2,2)}/2$ at a given time $t$ during the inspiral. Consider how this scaling affects the frequency content of the individual modes at the beginning of waveform generation. When a user requests a starting frequency of a waveform, e.g., $f_{\rm start}^{(22)}=20\ \mathrm{Hz}$, this corresponds to $f_{\rm start}^{(33)} \simeq 30\ \mathrm{Hz}$, and $f_{\rm start}^{(21)} \simeq 10\ \mathrm{Hz}$ for the (3,3) and (2,1)-modes respectively. An important detail is then that the frequency-domain waveform of the (3,3)-mode therefore should \emph{not} contain any content between 20 and 30 Hz\footnote{This can be seen from Eq.~\eqref{eq:SPA}, as the (3,3)-mode had an instantaneous frequency between 20 and 30 Hz only \emph{before} the (2,2)-mode was in the band of our detector. The (2,1)-mode on the other hand \emph{should} contain frequency content between 10 and 20 Hz, but because it is usually implied that a PE analysis is performed starting at $f_{\rm min}=f^{(2,2)}_{\rm start}$, this can be ignored.}.

As it is our goal to reproduce the time-domain model, where the FFT is used to compute the Fourier domain waveform which has a given duration $T$, the FFT will also only contain frequencies above $f_{\rm start}^{(\ell,m)}$ for the individual modes. Note that this is a difference to the frequency-domain \texttt{IMRPhenom} family of waveform models, which uses the SPA of PN equations for the inspiral part of the waveform, and where the same starting frequency is used for all modes (i.e., the $m=3$ modes also contain frequency component between 20 and 30 Hz when $f_{\rm start}$ is set to 20 Hz).

In the \vfive\ models, one instead employs a conditioning of the time-domain polarizations when generating a frequency-domain waveform. This involves starting the waveform at a starting frequency which is a bit below the requested starting frequency of the (2,2)-mode, and then applying a Hanning window
\begin{equation} \label{eq:Hann}
    H(t) = \frac{1}{2}\left[1-\cos \left(\pi \frac{t-t_0}{t_{\rm taper}}\right)\right]\ ,
\end{equation}
to the first few cycles of the waveform. This ensures that the Fourier transform of the waveform does not contain Fourier artefacts due to spectral leakage, but also that the Fourier transform does not contain unphysical low-frequency content in the $m>2$ modes. 

For the \vfivethmfd\ model, we want to recreate this and therefore apply a similar conditioning to the frequency-domain modes instead of the time-domain polarizations. To this end we write our time-domain, Hanning-tapered waveform modes as
\begin{equation} \label{eq:time_domain_tapered}
    h_{\ell m}^{\rm tapered}(t) =  \frac{a_0}{2}\left[1-\cos \left(\pi \frac{\Delta t}{t_{\rm taper}}\right)\right] e^{-i \Phi (t)}\ ,
\end{equation}
where $t_{\rm taper}$ is the duration of the tapering, $a_0$ is the starting amplitude, and $\Phi (t)$ is a quadratic approximation of the phase of the tapered waveform. We explain in more detail in Appendix~\ref{app:conditioning} how one can use this with the SPA from Eq.~\eqref{eq:SPA} to derive a tapering in the frequency domain that is equivalent to the approach in the time domain. We apply the resulting frequency-domain tapering to the $m>2$ SPA waveform modes, while just starting the $m\leq 2$ modes at the requested (2,2)-mode starting frequency instead.

\subsubsection{Evaluating the entire SPA waveform}

Having determined which part of the time-domain waveform to use for the SPA, we can compute the necessary ingredients for Eqs.~\eqref{eq:SPA} and \eqref{eq:SPA_components}. To this end, we  need to evaluate up to second time derivatives of the phases $\ddot{\phi}_{\ell m}(t)$. This can become problematic as the individual waveform cycles may not be resolved on the sparse grid of the dynamics. This in turn can lead to jumps by multiples of $2\pi$ in the phase of the different modes, which hinders the fitting to splines. To circumvent this issue, we therefore employ the idea of a carrier signal as outlined in Ref.~\cite{Cotesta:2020qhw} and already employed in the time-domain version of the \texttt{SEOBNRv5} models~\cite{Pompiliv5}.

Recall that, due to Eq.~\eqref{eq:m_scaling}, a mode with magnetic number $m$ will behave as $e^{-im\phi_{\rm orb}}$. Given our knowledge of the orbital phase from our solution of the EOB dynamics, we can therefore construct a slowly varying function by multiplying the mode by $e^{+im\phi_{\rm orb}}$ as
\begin{equation}\label{eq:deforbitalphasing}
	h_{\ell m}^{\rm carrier} = h_{\ell m} e^{+i m \phi_{\rm orb}} \simeq \left| h_{\ell m} \right| e^{i\, 0} \,,
\end{equation}
which will oscillate very slowly and can easily be fitted to cubic splines. The orbital phase is furthermore unwrapped in the dynamics (i.e., not constrained to be within $[-\pi,\pi]$ or similar intervals, so it is not an issue that not every orbital cycle is resolved). From Eq.~\eqref{eq:deforbitalphasing}, the higher derivatives of the phase can then be computed as
\begin{equation}\label{eq:derivatives_r}
	\frac{\d^n }{\d t^n} \phi_{\ell m} = \frac{\d^n }{\d t^n} \phi_{\ell m}^{\rm carrier} - m \frac{\d^n }{\d t^n} \phi_{\rm orb} \,.
\end{equation}

A final point that needs to be addressed is the potential zero-crossings of the amplitudes $|h_{\ell m}|=0$ of some of the modes, which is a prediction of PN theory for certain spinning systems. If one requires the amplitudes to be larger than zero, such a zero-crossing of the amplitude will lead to phase-jumps by $\pi$ which breaks the continuity of the phase and therefore our ability to take derivatives of it. One can however circumvent this issue by allowing the mode amplitudes to become negative instead of naively taking absolute values of the mode amplitudes\footnote{To this end, we check for increases of the carrier phase $\phi_{\ell m}^{\rm carrier}$ by close to $\pi$ and look for discontinuities in the phase if that is the case. If so, we multiply the amplitude by -1 from this time onwards, while adding a $\pi$-shift to the phase after the zero-crossing, which ensures that the phase is continuous and can be fitted to splines.}.

With these ingredients, one can use cubic splines to compute Eqs.~\eqref{eq:SPA} and \eqref{eq:SPA_components} on flexible frequency grids in a robust and accurate way. We note that the SPA for the \texttt{TEOBResumSPA} model is instead using a fourth-order finite difference formula for non-uniform grids using Lagrangian interpolants (see the Supplemental Material of Ref.~\cite{Gamba:2020ljo}). We find that our approach of using the time-domain cubic splines used for the waveform interpolations is already more robust and accurate, which can be easily seen in Fig.~\ref{fig:comparison}. There, we observe a large amount of numerical noise in both the Fourier phase and amplitude in the \texttt{TEOBResumSPA} waveform model in contrast to our \vfivethmfd\ implementation.

\subsection{The FFT for the late and post-inspiral} \label{subsec:FFT}

The next step to build our frequency-domain model is to compute the late- and post-inspiral-part of the waveform in the frequency domain via an FFT, which we intend to interpolate in the frequency domain to evaluate it on flexible frequency grids. Let us start by reviewing the FFT and relevant quantities for this algorithm. 

The FFT computes the Fourier transform of some discrete, equally-spaced, time-domain waveform data $\bm{h}=\left(h_0, h_1, \dots, h_{N-1}\right)^T$. Assuming a convention as in Eq.~\eqref{eq:Fourier_transform}, our sampled data is evaluated as $h_m = h(m\dt)$, defined on the interval $[0,\ T_{\rm FFT}]$. We further discuss the relationship between the interval $[0,T]$ of the SPA approach and $[0,\ T_{\rm FFT}]$ in Sec.~\ref{subsec:combination}. The number of time samples $N$ depends on the sampling rate $f_s = 1/\Delta t$, as $N = T_{\rm FFT} / \Delta t$.

The FFT then computes the discrete Fourier components of our data via
\begin{equation} \label{eq:FFT}
    \tilde{h}_k = \tilde{h}(f_k) =  \dt \sum_{m=0}^{N-1} h_m e^{-2 \pi i k m / N}\,
\end{equation}
at the frequency of the $k$-th bin $f_k = k / T_{\rm FFT}$. The frequency resolution $\Delta f= 1/T_{\rm FFT}$ therefore depends on the signal duration and can be increased via zero-padding of the data. The maximum frequency of our FFT signal is the Nyquist frequency (or maximum frequency), related to the sampling rate $f_s$ as
\begin{equation} \label{eq:fmax}
    f_{\rm max}=f_{\rm Nyquist}=1/(2\dt)=f_s/2\ .
\end{equation} 
Higher frequencies in the signal can not be resolved, but can influence the lower-frequency components of our signal due to aliasing, as we discuss in Sec.~\ref{subsec:mismatches}. For $\dt \rightarrow 0$, the FFT from Eq.~\eqref{eq:FFT} approaches the Fourier transform of Eq.~\eqref{eq:Fourier_transform} of the function we have sampled from.

It is further useful to highlight the difference to the real-valued FFT, which is a variant of the complex-valued FFT. We remind the reader that the Fourier transform and complex conjugation do not commute, but rather obey the symmetry relation
\begin{equation} \label{eq:symmetry_property}
    \widetilde{h^*}(f) = \tilde{h}^*(-f) \,,
\end{equation}
where the Fourier transform of the conjugated signal $h^*$ is the complex conjugate of the Fourier transformed signal $\tilde{h}$ evaluated at the negative frequency.

Assuming real-valued input data, one therefore only needs to compute the Fourier components for positive frequencies, as the negative frequencies are just the complex conjugate of the positive frequencies $\tilde{h}(f) = \tilde{h}^*(-f)$. This is in general not the case for complex signals, where one needs both the positive and negative frequencies to reconstruct the original time-domain signal loss-free. We want to highlight that our waveform modes $m\lessgtr 0$ only contain frequency content $f\gtrless 0$ due to the monotonicity and sign of the phase evolution.

The real-valued FFT is more efficient, as it only computes half the number of frequency-components. However, it can not be used for the individual modes of our waveform, as they are complex-valued. We therefore use the complex FFT for our waveform modes, and disregard the positive frequencies in the end, as they can be assumed to be zero due to the monotonicity of the phase of the modes (see Sec.~\ref{subsec:SPA}).

Assuming that we have already decided on a transition time $\ttrans$ in the previous step of Eq.~\eqref{eq:ttrans}, we interpolate all waveform modes from a starting time 
\begin{equation}
    t_{\rm start}^{\rm FFT} = \min_{\ellm} \left\{ \ttrans^{\ellm} - 3/\sqrt{\dot\omega_{\ell m}(t_{\rm trans})}-2000\ M \right\}\ 
\end{equation}
onwards. Stepping back from the transition time is a necessity to faithfully resolve the Fourier transform at the transition frequency, given our discussion around Eq.~\eqref{eq:support}. For a best-effort Fourier waveform, it is also important to reduce the effect of spectral leakage. To this end, we taper the first $2000\ M$ of our waveform with a Hanning tapering as in Eq.~\eqref{eq:Hann}, which is also the reason why we step back an additional $2000\ M$ from the end of the SPA.

The resulting frequency-domain polarizations are then interpolated via cubic splines such that they can be evaluated on a flexible frequency grid for use in multibanding and relative binning.

\subsection{Combining the methods} \label{subsec:combination}

In order to combine the SPA from Sec.~\ref{subsec:SPA} with an FFT of the late-inspiral-merger waveform from Sec.~\ref{subsec:FFT}, we finally need to stitch the modes from the SPA and FFT approaches together. For this combination we have three possibilities, given that our signal has monotonically evolving frequency. We can either choose a sharp transition time, a sharp transition frequency, or both, as we explain below. 

Assume that we want to evaluate our signal at (not necessarily equidistant) frequencies $\mathbf{f}=(f_1, f_2, \dots, f_N)$, which lie between $-f_{\rm max}$ and $f_{\rm max}$. We can also assume that our signal has only positive or negative frequency components dependent on the sign of the magnetic number $m$, as per our discussion at the beginning of Sec.~\ref{subsec:SPA}.

For a signal with a monotonically evolving instantaneous frequency, we then want to combine our methods such that
\begin{align}
    \label{eq:h_td}
    h_{\ell m}(t) = \left\{
            \begin{array}{ll}
                h^{\rm SPA}_{\ell m}(t), & \quad t < t_{\rm trans}^{\ellm}, \\\\
               h^{\rm FFT}_{\ell m}(t), & \quad t \geq t_{\rm trans}^{\ellm}.
            \end{array}
        \right. 
    \end{align}
The possibilities for the combination of the two methods are then as follows.

\paragraph{Combining both approaches at a sharp transition-time.} The most straightforward way to combine the two approaches is to simply Fourier transform both time-domain signals from Eq.~\eqref{eq:h_td} separately, assuming a discontinuous step to (from) 0 at $t_{\rm trans}^{\ellm}$ for the SPA (FFT) signal, and to then add the resulting frequency-domain waveforms together. This results in two frequency-domain waveforms $\tilde{h}^{\rm (SPA)}_{\ell m}(f)$ and $\tilde{h}^{\rm (FFT)}_{\ell m}(f)$ that, due to the linearity of the Fourier transform, can just be added across all frequencies to result in the correct frequency-domain waveform. As we have a discontinuity at $t_{\rm trans}^{\ellm}$, both frequency-domain waveforms will however have a non-vanishing contribution above and below the corresponding $f_{\rm trans}^{\ellm}$ due to the rectangular windowing, so we need to evaluate both the SPA and FFT waveform at all frequencies and add the two contributions together. For the SPA part of the signal, this would be equivalent to extending our SPA description of Sec.~\ref{subsec:SPA} to include the case of discontinuous edges, which leads to Fresnel integrals due to the limits in the integral of Eq.~\eqref{eq:gaussian}. As we furthermore have to evaluate both the SPA and FFT waveforms at more frequencies, and as we would have to change the SPA implementation, we have not pursued this approach further. It might however be interesting to explore in the future if one wants to change our mode-by-mode approach to an approach only applied to the polarizations.

\paragraph{Combining both approaches at a sharp transition time and frequency.} Another approach is to simply use the SPA waveform for frequencies below the transition frequency, and the FFT waveform for frequencies above the transition frequency. This can be easily implemented by evaluating the SPA and FFT waveforms up to the transition frequency $f_{\rm trans}^{\ellm}$ and evaluating the FFT waveform above the transition frequency.
\begin{align}
    \label{eq:h_fd}
    \tilde{h}_{\ell m}(f) = \left\{
            \begin{array}{ll}
                \tilde{h}^{\rm (SPA)}_{\ell m}(f), & \quad f < f_{\rm trans}^{\ellm}, \\\\
                \tilde{h}^{\rm (FFT)}_{\ell m}(f), & \quad f \geq f_{\rm trans}^{\ellm}
            \end{array}
        \right. \ .
\end{align}
We note that this way of combining the two approaches is not continuous at the transition frequency, as the SPA and FFT waveforms are not guaranteed to have the same amplitude and phase at the transition frequency, as can also be seen in Fig.~\ref{fig:comparison}. We have however been careful to assume the signal continuing indefinitely in Eq.~\eqref{eq:ttrans} and Eq.~\eqref{eq:h_td}, such that we can use the version of Eq.~\eqref{eq:gaussian} with infinite integration limits. By assuming that there is sufficient signal before and after the transition time, both the SPA and FFT waveforms should still agree to a reasonable level at the transition frequency. Furthermore, the discontinuity is not an issue for our applications, as it introduces negligible mismatches, as we show in Sec.~\ref{subsec:mismatches}. Conceptually, an accumulated error due to the SPA being inaccurate should furthermore produce worse results than a slight percent level change in amplitude and phase at the one instance in time where we combine SPA and FFT. We therefore choose this approach for our \vfivethmfd\ model.

\paragraph{Combining both approaches at a sharp transition-frequency.} A last possibility, which we have not pursued, is to use a sharp transition frequency and evaluate all modes up to there, at very different transition times. An issue with this approach is the inclusion of modes with high enough magnetic numbers $m$. These modes reach the transition frequency already during the early inspiral, and one therefore needs to interpolate a lot of this mode onto the finely sampled grid used for the FFT. This can also be seen in Fig.~\ref{fig:pedagogical}, where the (3,3)-mode reaches the (2,2)-mode transition-frequency much earlier than the other modes. As this would imply interpolating this mode much more, it would reduce the speed-gain of our approach, and we have therefore not pursued this approach further.

Having decided on approach \emph{b} to combine the different methods, we still have to account for the fact that the Fourier transform has an implicit time-dependence. If one shifts a signal in a Fourier transform as in Eq.~\eqref{eq:Fourier_transform} by some time-shift $\tau$, note that one acquires a frequency-dependant phase shift
\begin{equation} \label{eq:time_shift}
    \tilde{h}_\tau(f) = \int_{-\infty}^\infty \d t \, h(t-\tau)\, e^{-2i \pi f t} =  e^{-2i \pi f \tau} \tilde{h}(f) \ ,
\end{equation}
through substitution. Both our SPA and FFT signal need to have the same underlying time array, else they will differ by this frequency-dependent phase. The combination of the two signals at a sharp transition frequency of Eq.~\eqref{eq:h_fd} would therefore not produce Eq.~\eqref{eq:h_td}. Instead, we have to be conscious on defining both $h_{\ell m}^{\rm SPA}$ and $h_{\ell m}^{\rm FFT}$ on a common, implicit time grid. To this end, it is most convenient to align both signals at the common $\tmerge$, which is typically set to occur at $t=0$ in PE.

For the SPA signal, the interval $[0,\ t_{\rm trans}]$ already corresponds to the dynamics time. To change the resulting frequency-domain signal to have the merger at $t=0$, one therefore needs to apply a time shift by $\tau_{\rm SPA} = \tmerge$ to the frequency-domain waveform as in Eq.~\eqref{eq:time_shift}. The FFT is instead defined on the interval $[0,\ T_{\rm FFT}]$, which corresponds to the dynamics time $[t_{\rm start}^{\rm FFT},\ \tmerge + \Delta T]$, with $\Delta T$ the time after the merger in the interpolated waveform. The necessary time shift for alignment is therefore a different one and reads $\tau_{\rm FFT} = \tmerge - t_{\rm start}^{\rm FFT}$. 

The above expressions are applied straightforwardly to each mode
$h_{\ell m}$. In the final step, we will then combine the modes to obtain the two polarizations in the frequency domain
\begin{subequations} \label{eq:fd_polarizations}
    \begin{align}
    \tilde{h}_{+}(f) & = \frac{1}{2} \sum_{\substack{\ell\geq 2 \notag \\ \ell \geq m > 0}} \, \Big[ \,_{-2}Y_{\ell, m}^* + (-1)^{\ell} \, _{-2}Y_{\ell, -m} \Big] \\[-2.5ex] & \ \qquad \qquad \qquad \ \times \tilde{h}^*_{\ell, m}(-f)\ , \\
    \tilde{h}_{\times}(f) & =-\frac{i}{2} \sum_{\substack{\ell\geq 2 \\ \ell \geq m > 0}} \, \Big[ \,_{-2}Y_{\ell, m}^* + (-1)^{\ell+1} \, _{-2}Y_{\ell, -m} \Big] \notag \\[-2.5ex] & \ \qquad \qquad \qquad \  \times \tilde{h}^*_{\ell, m}(-f)\ ,
    \end{align}
\end{subequations}
where $f>0$ is assumed. We derive these expressions in Appendix~\ref{app:polarizations}. Note that the Fourier transform of the modes is evaluated at negative frequencies and only for $m>0$ modes, which is a consequence of our sign conventions, as we explain in Sec.~\ref{subsec:SPA}. In contrast to the time-domain version of the model, we therefore do not compute the $m<0$ modes at all, which also contributes to a slight speed-up for the evaluation of the waveform modes.

As a final note, the \vfive\ models obey the convention that the primary body has mass $m_1 \geq m_2$ and the reference orbital phase is defined with respect to the primary body, which implies a $\pi$ rotation of the system if the user instead specifies a system with $m_1 < m_2$. For our model, this is implemented by a simple $\pi$-shift in the orbital phase $\phi_{\rm ref} \mapsto \phi_{\rm ref} + \pi$, instead of a phase shift of the modes as $h_{\ell m} \mapsto h_{\ell m} e^{i m \pi} = (-1)^m h_{\ell m}$, as is implemented in the time-domain version of the \vfive\ models.

\section{Application in multibanding and relative binning} \label{sec:mb_rb}

The frequency-domain waveform model developed in the previous section accelerates both waveform generation and likelihood evaluation compared to the time-domain version of the \vfivethm\ model. The speed-up arises due to two effects. First, during the inspiral the SPA replaces the construction of a finely sampled time-domain signal followed by a long FFT with a direct evaluation of the Fourier-domain polarizations, which is particularly costly for BNS systems.

Second, a frequency-domain model can be evaluated directly on flexible, non-uniform frequency-grids. For long BNS signals starting at low frequencies, the number of equidistant frequency bins can reach $\mathcal{O}(10^6$--$10^8)$, rendering repeated likelihood evaluations during stochastic sampling computationally expensive. Multibanding and relative binning likelihoods on the other hand approximate the likelihood through some assumptions and can reduce the necessary frequency bins to $\mathcal{O}(10^2$--$10^4)$ while retaining the necessary amount of accuracy for PE. Due to their derivation in the frequency domain, they are however not naturally compatible with time-domain models, which rely on an equidistant FFT sampling.

It has to be noted that during the development of this work, a version of the heterodyned likelihood has been developed in Ref.~\cite{Sharma:2026wqw} which can be used immediately in the time domain. The introduced method combines ideas of both multibanding and relative binning in the time domain and can be explored as an alternative to the methods described in this work in the future. The advantage of our approach is however, that one can immediately use the thoroughly tested likelihood and sampling as implemented in \texttt{Bilby}~\cite{Ashton:2018jfp}. 

In this section, we therefore focus on the application of our frequency-domain model within standard multibanding and relative binning likelihoods.
To this end, we review the conventional likelihood used in PE in Sec.~\ref{subsec:likelihood}, and the changes multibanding and relative binning introduce in Secs.~\ref{subsec:multibanding} and \ref{subsec:relbin}, respectively. We suggest readers who are primarily interested in the original results of this work to proceed directly to Sec.~\ref{sec:speed_robustness}, where we demonstrate the resulting speed improvements and robustness assessments.

\subsection{The full likelihood} \label{subsec:likelihood}

Parameter estimation for compact binary coalescences is performed in a Bayesian framework. The posterior distribution for source parameters $\boldsymbol{\theta}$ given data $d$ is
\begin{equation}
    p(\boldsymbol{\theta}\mid d)
    \propto
    p(d\mid\boldsymbol{\theta})\,p(\boldsymbol{\theta}) .
\end{equation}
Assuming stationary, Gaussian detector noise, the likelihood for a single detector can be written in terms of the noise-weighted inner product~\cite{Thrane:2018qnx}
\begin{equation} \label{eq:innerprod}
(a|b) = 4 \, \mathrm{Re} \sum_k \frac{a^*(f_k)b(f_k)}{S_n(f_k)} \Delta f ,
\end{equation}
where $S_n(f)$ is the one-sided noise power spectral density.

The log-likelihood then takes the compact form
\begin{equation}
\log p(d\mid\boldsymbol{\theta}) = -\frac{1}{2} (d-h \,|\, d-h) + \text{const.}
\label{eq:loglike_innerprod}
\end{equation}
which can be expanded to read
\begin{equation}
\log p(d\mid\boldsymbol{\theta}) = -\frac{1}{2} \left[ (d|d) - 2\ \mathrm{Re}(d|h) + (h|h) \right] + \text{const.}
\label{eq:loglike_expanded}
\end{equation}
Both multibanding and relative binning can be understood as controlled approximations to the inner products appearing in Eq.~\eqref{eq:loglike_expanded}, while leaving the overall Gaussian likelihood structure unchanged. We briefly review these methods to emphasize why the arbitrary-frequency evaluation in our frequency-domain waveform model is necessary, as we will see in Sec.~\ref{sec:pe}.

\subsection{Multibanding}  \label{subsec:multibanding}

The multibanding method~\cite{Vinciguerra_2017,Morisaki:2021ngj}, closely related to the heterodyned likelihood~\cite{Cornish:2021lje}, reduces the computational cost of likelihood evaluation by replacing the dense, uniformly spaced frequency grid with a set of frequency bands of varying width. It has been implemented in inference frameworks such as \texttt{Bilby}~\cite{Ashton:2018jfp} and \texttt{LALInference}, yielding order-of-magnitude speedups for long BNS signals with negligible bias when properly configured.

The key observation is that for chirping compact binaries the time-to-merger at lower frequencies is much longer than at higher frequencies. Consequently, one can use a coarser frequency resolution at higher frequencies than at lower frequencies without introducing significant errors.

This idea can be used iteratively by partitioning the frequency-domain waveform into bands $\mathcal{B}^{\rm MB}_j = [f_j^{\rm min}, f_j^{\rm max}]$ at different frequency resolutions. The inner product of Eq.~\eqref{eq:innerprod} can then be approximated as
\begin{equation} \label{eq:mb_likelihood}
    (a|b)_{\rm MB}
    \simeq
    4 \, \mathrm{Re}
    \sum_{j}
    \sum_{m=1}^{N_j}
    w_{jm}
    \frac{
        a^*(f_{jm}) b(f_{jm})
    }{
        S_n(f_{jm})
    },
\end{equation}
where $f_{jm}$ are representative frequencies in band $j$ (non-uniform across the signal), $N_j$ is the reduced number of samples in that band, and $w_{jm}$ are weights chosen to approximate the full sum.

The multibanded likelihood is then obtained by replacing the inner products in Eq.~\eqref{eq:loglike_expanded} with their multibanded version, apart from the data-only term $(d|d)$, which can be computed once, and treated as constant.

\begin{figure*}[t]
    \centering
   \includegraphics[width=0.99\linewidth]{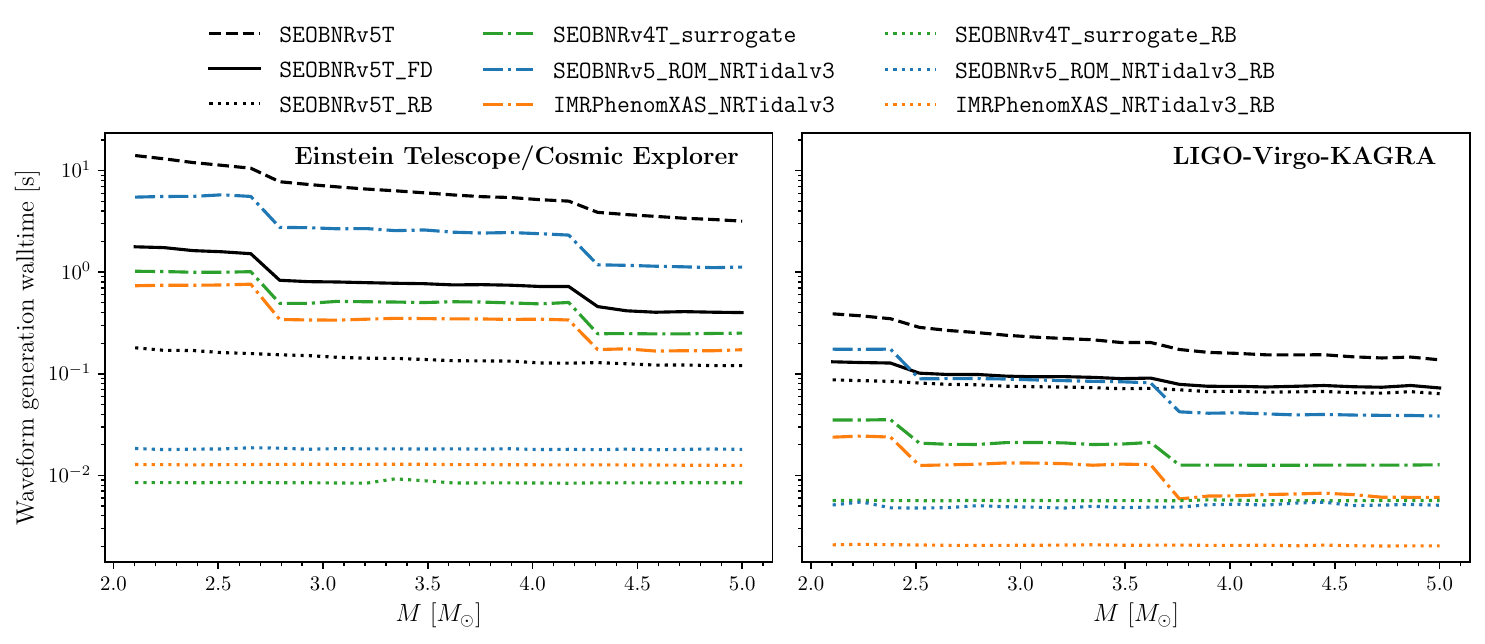}
   \caption{Waveform generation time comparisons of \vfivethm\ and \vfivethmfd\ (black) against other waveform approximants and the conventional FFT method. The left panel shows the timing for systems starting at 5 Hz in ET/CE, while the right panel shows the timings for systems starting at 20 Hz in LVK detectors. We compare waveforms evaluated on a full, uniform frequency grid (dashed, solid, and dash-dotted) to frequency-domain waveforms evaluated only on a relative binning frequency grid (with suffix \texttt{RB}, dotted). Note that an additional overhead of projecting the polarizations into the detector during PE is not included here.}
   \label{fig:benchmark}
\end{figure*}

\subsection{Relative binning} \label{subsec:relbin}

Relative binning~\cite{Zackay:2018qdy,Krishna:2023bug} accelerates likelihood evaluation by exploiting the similarity between nearby waveforms. The key idea is to approximate $(d|h)$ and $(h|h)$ in Eq.~\eqref{eq:loglike_expanded} by expressing the waveform in terms of a fiducial waveform $h_0(f)$ and a slowly varying ratio to nearby waveforms.
Defining the waveform ratio
\begin{equation}
    r(f;\boldsymbol{\theta})
    =
    \frac{h(f;\boldsymbol{\theta})}{h_0(f)},
\end{equation}
we express a waveform with parameters $\boldsymbol{\theta}$ as
\begin{equation}
    h(f;\boldsymbol{\theta})
    =
    h_0(f)\, r(f;\boldsymbol{\theta}) .
\end{equation}
For parameters $\boldsymbol{\theta}$ close to the true (or fiducial) parameters $\boldsymbol{\theta}_{0}$, the ratio $r(f;\boldsymbol{\theta})$ is expected to exhibit significantly less structure than the original waveform $h(f;\boldsymbol{\theta})$, and to therefore vary slowly with frequency. The frequency domain is thus partitioned into bins $\mathcal{B}^{\rm RB}_j$, and $r(f)$ is approximated linearly within each bin,
\begin{equation} \label{eq:linear_ratio}
    r(f)
    \approx
    r_j^{(0)}
    + r_j^{(1)} (f - f_j^{\rm c}),
    \qquad
    f \in \mathcal{B}_j ,
\end{equation}
where $f_j^{\rm c}$ is a representative frequency of bin $j$. The frequency bins are furthermore determined based on a PN argument quantifying the expected change in waveform that is expected by a small variation of the parameters $\boldsymbol{\theta}$. This involves the definition of two tunable parameters, $\chi$ and $\epsilon$, that determine the expected change of PN coefficients and the allowed de-phasing per frequency bin in radian, respectively~\cite{Zackay:2018qdy}. They need to be chosen by the user of the relative binning method in addition to the parameters of the fiducial waveform before PE.

Applying this approximation to the likelihood of Eq.~\eqref{eq:loglike_expanded}, the data-only term $(d|d)$ remains unchanged, while the cross term and waveform-only term reduce to
\begin{equation} \label{eq:rb_likelihood}
    \log p(d\mid\boldsymbol{\theta})
    \simeq
    -\frac{1}{2}
    \left[
        (d|d)
        - 2\, \mathrm{Re} \sum_j A_j r_j
        + \sum_j B_j |r_j|^2
    \right],
\end{equation}
where $r_j$ denotes the interpolated ratio evaluated at the representative frequency of bin $j$. The number of such frequency bins is typically of order a few hundred and is largely independent of the total signal duration. The data-dependent quantities $A_j$ and $B_j$, involving the fiducial waveform, are precomputed once at the beginning of the inference run and read
\begin{equation}
    \begin{aligned}
    A_j & = 4 \sum_{f_k \in \mathcal{B}_j}
    \frac{d^*(f_k) h_0(f_k)}{S_n(f_k)} \Delta f,
    \\ 
    B_j & = 4 \sum_{f_k \in \mathcal{B}_j}
    \frac{|h_0(f_k)|^2}{S_n(f_k)} \Delta f .
    \end{aligned}
\end{equation}

The computational gain arises because the slowly varying ratio $r(f)$, and hence the waveform $h(f;\boldsymbol{\theta})$, needs to be evaluated only on the sparse, non-uniform set of representative frequencies associated with the bins $\mathcal{B}^{\rm RB}_j$. The relative-binning method has also been applied with the \texttt{TEOBResumSPA} model in Ref.~\cite{Huez:2025gja}. Extensions to higher harmonics and precessing systems perform the binning mode-by-mode in order to maintain accuracy (see, e.g., Ref.~\cite{Leslie:2021ssu}).

\section{Speed-improvements and robustness} \label{sec:speed_robustness}

In this section, we assess how constructing our frequency-domain implementation of the \vfivethm\ model directly in the frequency domain affects waveform-generation speed and PE for BNS signals. To this end, we assess its waveform generation speed-improvements in Sec.~\ref{subsec:benchmarks}. We further study the mismatch between our frequency-domain implementation and the time-domain version of the \vfivethm\ model in Sec.~\ref{subsec:mismatches}, where we also assess the implications for PE, and compare our results to other sources of systematic errors.

\subsection{Benchmarking and waveform generation performance} \label{subsec:benchmarks}

Figure~\ref{fig:benchmark} summarizes the waveform generation cost of our frequency-domain EOB implementation as a function of total mass for a representative BNS system. We compare the time-domain \texttt{SEOBNRv5THM} model to its frequency-domain implementation developed in this work, \texttt{SEOBNRv5THM\_FD}, and to established native frequency-domain models: \texttt{SEOBNRv4T\_surrogate}~\cite{Lackey:2018zvw}, \texttt{SEOBNRv5\_ROM\_NRTidalv3}, and \texttt{IMRPhenomXAS\_NRTidalv3}~\cite{Abac:NRT}. For comparability, we call all approximants only including the (2,2)-mode, as both \texttt{SEOBNRv4T\_surrogate} and the current \texttt{lal}-implementation of \texttt{NRTidalv3} are (2,2)-mode only approximants.

All walltimes correspond to full waveform generation on a single CPU core (Apple M2). We consider signals in LVK detectors and next-generation observatories (ET/CE). For LVK detectors, waveforms are generated between $f_{\rm min}=20$~Hz and $f_{\rm max}=2048$~Hz; for third-generation detectors, between 5~Hz and 2048~Hz as those detectors are expected to have a suppressed seismic and thermal noise contribution at low frequencies in the PSD, which will enable signal analyses starting at earlier times~\cite{Punturo:2010zz,Reitze:2019iox,Evans:2021gyd}.

We simulate a BNS configuration with mass ratio $q=1.4$, spins $\chi_1=\chi_2=0.1$, and tidal deformabilities $\Lambda_1=800$, $\Lambda_2=400$. The total mass is varied between $2\,M_\odot$ and $5\,M_\odot$ to probe different signal durations. The signal duration $T$ is taken from the corresponding time-domain \texttt{SEOBNRv5THM} waveform and determines the frequency resolution $\Delta f = 1/T$ of the full frequency-grid. As customary in LVK analyses, $T$ is padded to the next power of two to optimize FFT performance.

We compare the time-domain \texttt{SEOBNRv5THM} model (dashed lines), its direct frequency-domain implementation \texttt{SEOBNRv5THM\_FD} (solid lines), and native frequency-domain waveform models for BNS (dash-dotted lines) on the full-frequency grid as determined by the signal duration. We also include our frequency-domain EOB model evaluated only on the sparse frequency grid determined by the relative binning method as \texttt{SEOBNRv5THM\_RB} (dotted lines), where we use the relative binning parameters as we used for the PE in Sec.~\ref{sec:pe}, $\chi=1$ and $\epsilon=0.0025$. For ET/CE detectors, we reduce this even further to $\epsilon_{3\mathrm{G}}=0.00025$ as we expect the hour-long and high-SNR signals will need more data points for unbiased results than the minute long signals in current detectors. We do not include the multibanding frequency bins in this comparison, as the amount of frequency points should be comparable between the two methods and as the precise location of the frequencies to be evaluated should not affect our results.

Across the full BNS mass range, we find that the frequency-domain EOB model reduces waveform generation time by a factor between $2$ and $10$ compared to the time-domain EOB model. The speed-up is largest for low total masses, where signal durations are longest and the cost of time-domain evolution and Fourier transformation is highest. Although we do not show the walltimes when including higher-order modes, comparison shows that the speed-up is largest when including only the (2,2)-mode and slightly reduced with an increased number of higher modes, as our mode-by-mode construction of the frequency-domain polarizations requires more FFTs than the conventional time-domain waveform model. In general the waveform generation scales better than linearly with the number of modes.

At higher total masses, the runtime of the frequency-domain EOB model approaches a floor of $\sim 0.1\,s$, corresponding to the cost of integrating the underlying EOB dynamics combined with the cost of interpolating and FFT-ing the post-inspiral part of the waveform. Beyond this point, the runtime becomes effectively independent of signal duration. The relative binning variant isolates this irreducible cost most clearly, in particular in comparison to the full frequency-domain waveform.

\texttt{IMRPhenomXAS\_NRTidalv3} and \texttt{SEOBNRv4T\_surrogate} remain the fastest BNS waveform approximants overall, with typical walltimes between $0.01$ and $0.1\,s$. This reflects the fact that they do not solve the full EOB dynamics during waveform generation and are natively built in the frequency domain. Our frequency-domain EOB implementation therefore remains slower by a factor of a few to ten, reflecting the intrinsic cost of computing the physical EOB dynamics rather than interpolating a reduced-order representation.
Nevertheless, \texttt{SEOBNRv5THM\_FD} approaches and exceeds the performance of \texttt{SEOBNRv5\_ROM\_NRTidalv3} for total masses $M \lesssim 2\,M_\odot$ well within the astrophysically relevant BNS regime.

Signal durations of the systems considered here range from 128\,s to 512\,s in LVK detectors and from 4096\,s (1\,h) up to 16384\,s (4.5\,h) in third-generation detectors. For these long and data-intensive signals, the benefit of the direct frequency-domain construction becomes much more pronounced.
In particular, the frequency-domain EOB model reaches the $\sim 0.1\,s$ dynamics floor even for $4.5$-hour long signals when evaluating it only on the relative binning frequency bins, which corresponds to a speed-improvement by around one to two orders of magnitude compared to the time-domain model, effectively making PE for BNS in third-generation detectors with \texttt{SEOBNRv5THM} feasible. This also demonstrates that the waveform generation cost can become effectively decoupled from signal duration. The relative performance gap between our frequency-domain EOB model and native frequency-domain models over the full frequency grid also narrows to within a factor $\lesssim 2$ over most of the mass range.

The frequency-domain EOB construction presented here in summary substantially reduces the computational cost of tidal EOB waveforms while preserving the physical consistency and extensibility of the underlying model. Although native frequency-domain models can remain faster in absolute terms, the frequency-domain EOB implementation closes much of the performance gap, particularly in the astrophysically relevant BNS mass range and for long third-generation signals, while retaining the flexibility to incorporate future improvements to the EOB dynamics. 

When performing PE in Sec.~\ref{sec:pe}, note that these speed-improvements also immediately translate into our PE benchmarks in Table~\ref{tab:pe_runtime_comparison}, where we find speed-ups of around a factor of 2 for the full likelihood and up to a factor of 7 for the multibanding and relative binning likelihoods, which is consistent with the waveform generation speed-ups found here. It should be emphasized that this includes an additional speed-up during the likelihood evaluation, as Eqs.~\eqref{eq:mb_likelihood} and \eqref{eq:rb_likelihood} need to be evaluated on much fewer frequencies compared to the full frequency-grid, therefore reducing the amount of floating-point-operations per likelihood evaluation. Note for the full likelihood evaluation, that the amount of floating-point-operations in projecting the waveform into the detectors becomes the dominant cost. This can also be seen more evidently as the walltimes between a recovery using \texttt{SEOBNRv4T\_surrogate} and \texttt{SEOBNRv5THM\_FD} differ by less than a factor 2 when using the full likelihood.

\begin{figure}[t]
    \centering
   \includegraphics[width=0.99\linewidth]{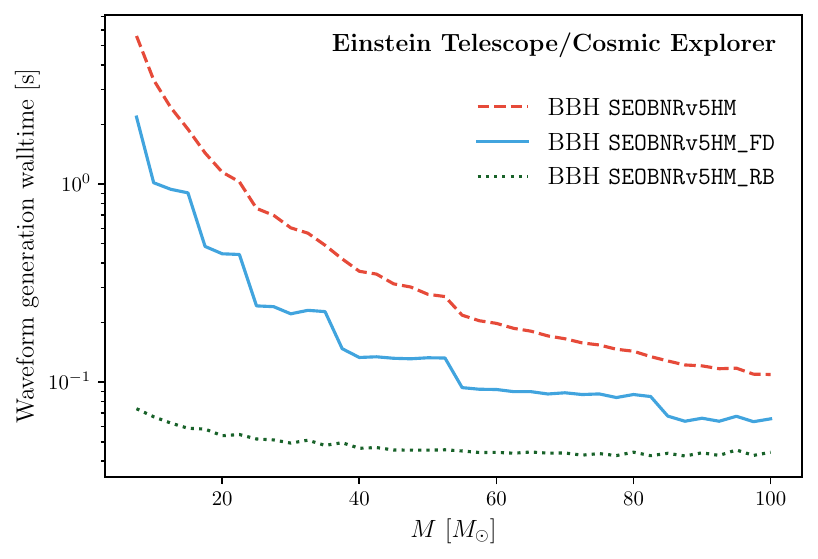}
   \caption{The walltimes for a $q=4$ BBH system in ET/CE using the BBH baseline model \vfivehm, starting at 5 Hz and including all higher-order modes. We compare the time-domain version of the model (red, dashed) to the frequency-domain version \texttt{SEOBNRv5HM\_FD} (blue, solid) evaluated on the full frequency-grid, and when evaluating it on the relative binning frequency bins (green, dotted, with suffix \texttt{RB}).}
   \label{fig:BBH_benchmark}
\end{figure}

As our approach described in the previous section is furthermore fully generic for spin-aligned, quasi-circular binaries, in Fig.~\ref{fig:BBH_benchmark}, we also show the speed-comparison when applying our approach to \vfivehm\ for BBH systems in ET detectors. Due to the shorter signal durations of BBH systems, the speed-up is less pronounced than for BNS systems, but we still find a factor of around 3 improvement in waveform generation time across the mass range when evaluating the full frequency-domain waveform. When evaluating only on the relative binning frequency bins, we find a speed-up of around an order of magnitude, which is consistent with the speed-up found for BNS systems when evaluating only on the relative binning frequency bins. Note that we sample all waveforms at 8 kHz and include all modes up to $\ell=5$ for this comparison, which is more costly than the BNS case where we only include the (2,2)-mode. For BBH systems in LVK detectors starting at 20 Hz, our approach does not make a difference at least above a total mass of $\sim 30\ M_{\odot}$, however, which is why we do not show results for this case.

\subsection{Studying different systematic error sources}
\label{subsec:mismatches}

The speed gains discussed in the previous section are only applicable if the frequency-domain implementation is sufficiently accurate and does not introduce biases in PE. We therefore quantify the systematic differences introduced by our frequency-domain construction by computing mismatches between different waveform realizations across the same region of the BNS parameter space.

To this end, we organize this section into three distinct categories of systematic effects in order to disentangle numerical implementation errors from genuine physical modeling uncertainties. 
\begin{enumerate}[label=(\roman*)]
    \itemsep-3pt
    \item We first assess implementation-level differences by comparing the frequency-domain model to its time-domain counterpart with and without the inclusion of higher-modes. 
    \item We then quantify physical modeling differences by comparing the time-domain approximants \texttt{SEOBNRv5THM} to \texttt{TEOBResumS-Dal\'{i}}. 
    \item Finally, we evaluate modeling and numerical approximations commonly employed in PE, such as neglecting subdominant higher-order modes and down-sampling the waveform. 
\end{enumerate}
This comparison allows us to place the frequency-domain-induced mismatches in the broader context of waveform systematics relevant for LVK analyses.

The GW signal emitted by a quasi-circular, aligned-spin BNS system depends on 13 parameters: the intrinsic parameters $\boldsymbol{\lambda}=\{m_{1,2},\chi_{1,2},\Lambda_{2}^{(1,2)}\}$, and the extrinsic  direction of the observer from the source $(\iota, \varphi)$, the luminosity distance $d_L$, the polarization angle $\psi$, 
the location in the sky of the detector $(\theta, \phi)$, and the coalescence time $t_c$. The detector strain can be written as
\begin{align}
\label{eq:det_strain}
h(t) =\,& F_+(\theta,\phi,\psi)\, h_+(t;\iota,\varphi,d_L,\boldsymbol{\lambda},t_c) \nonumber \\
&+ F_\times(\theta,\phi,\psi)\, h_\times(t;\iota,\varphi,d_L,\boldsymbol{\lambda},t_c),
\end{align}
where $F_{+,\times}$ are the antenna pattern functions~\cite{Sathyaprakash:1991mt,Finn:1992xs}. Equivalently, the strain can be expressed in terms of an effective polarization angle $\kappa(\theta,\phi,\psi)$~\cite{Cotesta:2018fcv}:
\begin{align}
\label{eq:strainKappa}
h(t)=\mathcal{A}(\theta,\phi)\bigl(h_+\cos\kappa+h_\times\sin\kappa\bigr),
\end{align}
where, for brevity, we suppress the functional dependences of $\kappa$, $h_+$, and $h_\times$.

To assess the agreement between two waveforms 
with higher-order multipoles~\cite{Cotesta:2018fcv,Ossokine:2020kjp,Garcia-Quiros:2020qpx}, which we denote as the signal observed by a detector, $h_s$, and the template, $h_t$, we define the faithfulness function (or match) \cite{Cotesta:2018fcv,Ossokine:2020kjp},
\begin{equation}
\mathcal{M}(\boldsymbol{\lambda},\iota_{s},{\varphi}_{s},\kappa_{s}) =  \max_{t_{c_t}, {\varphi}_{t}, \kappa_{t}}  \frac{ ( h_s|h_t )}{\sqrt{  ( h_s|h_s )  ( h_t|h_t )}},
\end{equation}
where the inner product is defined in Eq.~(\ref{eq:innerprod}) and the associated mismatch is 
\begin{equation} \label{eq:mismatch}
   \mathcal{M}\mathcal{M}(\boldsymbol{\lambda},\iota_{s},{\varphi}_{s},\kappa_{s}) = 1 - \mathcal{M}(\boldsymbol{\lambda},\iota_{s},{\varphi}_{s},\kappa_{s}).
\end{equation}

In the high-SNR limit, the mismatch is related to the indistinguishability criterion~\cite{Chatziioannou:2017tdw,Lindblom:2008cm}, which states that two waveforms are statistically indistinguishable if
\begin{equation} \label{eq:indistinguishability}
1-\mathcal{M}\lesssim \frac{D}{2\rho^2},
\end{equation}
where $\rho$ is the optimal SNR and $D$ the number of parameters not optimized over. For $\rho=60$ and $D=6$ ($D=8$ in the higher-mode case\footnote{As we explain below, we also fix the inclination $\iota_t=\iota_s$ and azimuthal angle $\varphi_t=\varphi_s$ in higher-modes case, while we optimize over the effective polarization angle $\kappa_t$. This implies two extra parameters in the indistinguishability condition as explained in Ref.~\cite{Thompson:2025hhc}.}), this corresponds to a targeted limit $1-\mathcal{M}\lesssim8\times10^{-4}$ ($1-\mathcal{M}\lesssim10^{-3}$) to avoid biases at the $1\sigma$ level in PE~\cite{Thompson:2025hhc}. Note that this condition is necessary but not sufficient for biases in PE.

Typically, we set the inclination angle of the template and the signal to be the same, 
while the coalescence time, azimuthal and effective polarization angles of the template, $(t_{c_t},\varphi_{t}, \kappa_t)$, 
are adjusted to maximize the faithfulness of the template.  The maximizations over the coalescence time $t_{c_t}$, 
and coalescence phase ${\varphi}_{t}$ are performed numerically, while the optimization over the effective 
polarization angle $\kappa_{t}$ is done analytically as described in Ref.~\cite{Capano:2013raa}.

For $(2,2)$-mode-only waveforms, differences in $\iota$, $\varphi$, and $\kappa$ are degenerate and can be absorbed into a global phase shift, so that optimizing over $(t_{c_t},\kappa_t)$ is sufficient to account for all three extrinsic angles. This well-known degeneracy no longer holds once higher-order modes are included. With higher modes, each multipole carries a distinct azimuthal dependence $e^{im\varphi}$ and inclination dependence through $_{-2}Y_{\ell m}(\iota,\varphi)$ in Eq.~\eqref{eq:hoft_sphericalH}, which breaks the simple phase degeneracy. 

In standard higher-mode analyses one therefore fixes the inclination and maximizes over $(t_{c_t},\varphi_t,\kappa_t)$~\cite{Cotesta:2018fcv,Ossokine:2020kjp,Pompiliv5,Gamboa:2024a}. Since we explicitly modify the computation of the frequency-domain polarizations in Eq.~\eqref{eq:fd_polarizations}, we instead enforce $\varphi_t=\varphi_s$ to test the internal consistency of the frequency-domain construction under identical conventions. We therefore optimize only over $(t_{c_t},\kappa_t)$ in the higher-mode case. Our reported mismatches when including higher-order-modes are conservative upper bounds, as any additional maximization over $\varphi_t$ can only decrease $\mathcal{M}$.

To reduce the dimensionality of the faithfulness function one usually defines the sky-and-polarization-averaged faithfulness \cite{Babak:2016tgq,Ossokine:2020kjp,Pompiliv5,Gamboa:2024a} as
\begin{equation} \label{eq:averaged_mismatch}
	\overline{\mathcal{M}}\left(\boldsymbol{\lambda}, \iota_{s}\right) = \frac{1}{8 \pi^2} \int_0^{2 \pi} \d \kappa_{s} \int_0^{2 \pi} \d \varphi_{s} \mathcal{M}\left(\boldsymbol{\lambda}, \iota_{s}, \varphi_{s}, \kappa_{s}\right).
\end{equation}
Rather than employing the sky-position-and-polarization-averaged faithfulness, we explicitly sample $\varphi_s\in[0,2\pi]$ and $\kappa_s\in[0,2\pi]$ instead.
Our rationale is that, while the averaged quantity reduces the dimensionality of our mismatches very slightly, it can also obscure regions in sky-position or polarization angles where mismatches are enhanced due to issues in the waveform model, which we would like to be sensitive to.
By retaining the full distribution over $(\varphi_s,\kappa_s)$, we therefore avoid averaging out potential tails, while the resulting sample mean over $(\varphi_s,\kappa_s)$ remains consistent with Eq.~\eqref{eq:averaged_mismatch}. We can also directly assess the dependence of the mismatch on sky position and effective polarization to identify any potential problems.

\begin{figure}[t]
    \centering
   \includegraphics[width=0.99\linewidth]{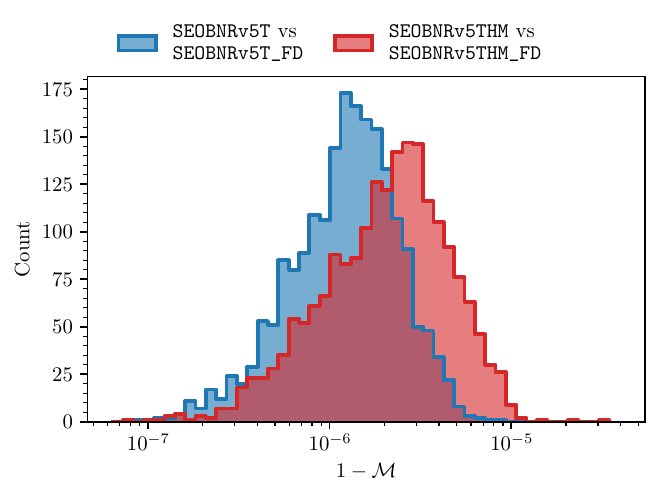}
   \caption{The mismatches across the BNS parameter-space between the \vfivethmfd\ and \vfivethm\ model focusing only on the dominant (2,2)-mode (blue) and when including also the (2,1) and (3,3)-mode (red).}
   \label{fig:mm_small}
\end{figure}

\begin{figure*}[t]
    \centering
   \includegraphics[width=0.99\linewidth]{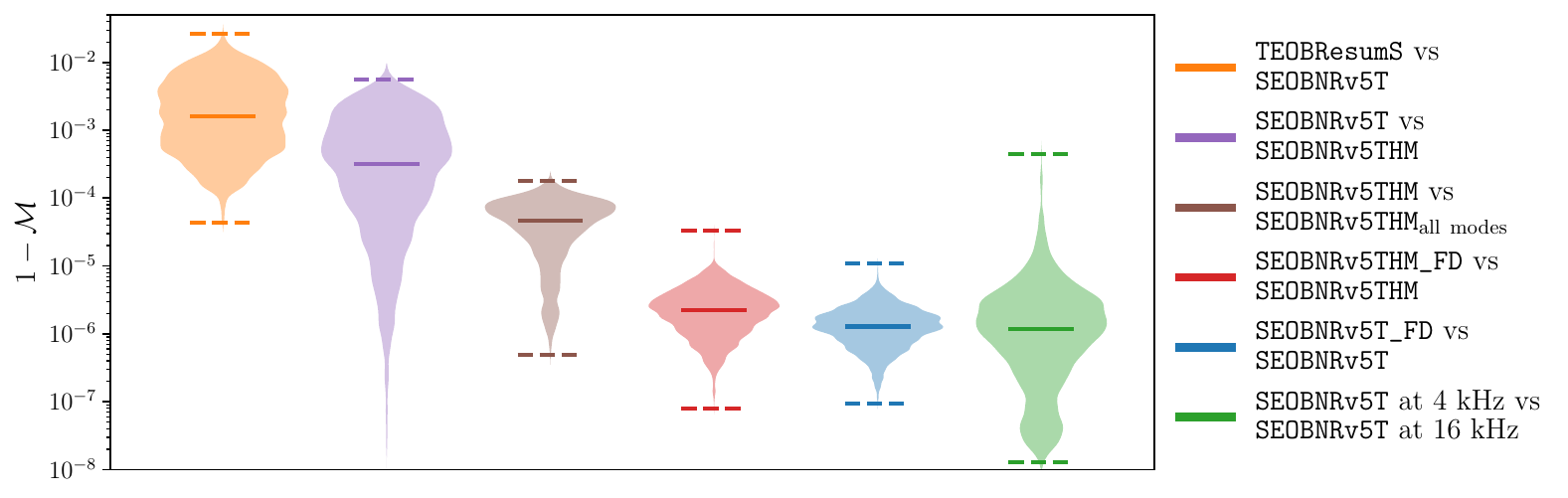}
   \caption{A comparison of mismatches between different models or approaches across the same BNS parameter space. We indicate minimum and maximum mismatch (dashed) and median mismatch (solid) for each distribution.}
   \label{fig:mm}
\end{figure*}

Throughout this section, we therefore sample binaries with $m_{1,2}\in[0.5,3]\,M_\odot$, $\chi_{1,2}\in[-0.5,0.5]$, and $\Lambda_{2}^{(1,2)}\in[0,5000]$. For higher-mode systems we additionally draw the extrinsic signal parameters $\varphi_s\in[0,2\pi]$, $\iota_s\in[-\pi,\pi]$ and $\kappa_s\in[0,2\pi]$. Mismatches are computed using Eq.~\eqref{eq:mismatch}, optimizing over $(t_{c_t},\kappa_t)$ for each system, which corresponds to an optimization over a global phase for (2,2)-mode-only studies. We compute all mismatches between 20 Hz and 2048 Hz and use the design target for O5 from Ref.~\cite{LVK_psds}. The mismatch distributions shown in Figs.~\ref{fig:mm_small} and \ref{fig:mm} are obtained from 2000 samples over this parameter space.

As in Sec.~\ref{subsec:benchmarks}, \texttt{SEOBNRv5T} denotes the time-domain model with only the $(2,2)$ mode, while \texttt{SEOBNRv5THM} additionally includes the $(2,1)$ and $(3,3)$ modes, and $\texttt{SEOBNRv5THM}_{\rm all\ modes}$ contains the (3,2), $(4,4)$, $(4,3)$, and $(5,5)$ mode, as well. 

We compare the frequency-domain implementations \texttt{SEOBNRv5THM\_FD} and \texttt{SEOBNRv5T\_FD} to their time-domain counterparts \texttt{SEOBNRv5THM} and \texttt{SEOBNRv5T}, while keeping the mode content consistent between the frequency- and time-domain implementations. The resulting mismatches are shown in Fig.~\ref{fig:mm_small}. Note that we call both approximants at a sampling frequency of 16 kHz to avoid any potential aliasing effects in the Fourier transform, which we will discuss in more detail below.

We find that with and without higher-modes, the mismatches remain between the $10^{-7}$ and $10^{-5}$ level with a tail of mismatches above the $10^{-5}$ level only when including higher-order modes. These mismatches are two to three orders of magnitude smaller than inter-model differences and much below the mismatch requirement of Eq.~\eqref{eq:indistinguishability}, demonstrating that the frequency-domain construction reproduces the time-domain baseline to the necessary numerical accuracy even when one considers the impact of higher-order modes in current and planned detectors. Note that when using Eq.~\eqref{eq:indistinguishability}, a mismatch of $10^{-5}$ corresponds to bias only at an SNR of $\mathcal{O}(500)$, which means our approach should remain valid even for ET and CE sources.

To contextualize the frequency-domain-induced mismatches, we compute $(2,2)$-mode mismatches between \texttt{SEOBNRv5T} and \texttt{TEOBResumS} as well as within-approximant mismatches under assumptions common in PE: (i) neglecting the $(2,1)$ and $(3,3)$ modes in a (2,2)-mode only analysis; (ii) neglecting the additional $(3,2)$, $(4,4)$, $(4,3)$, and $(5,5)$ modes; and (iii) evaluating the same model at different sampling rates (4\,kHz vs.\ 16\,kHz). The latter tests aliasing effects when down-sampling 16\,kHz data to 4\,kHz, which can introduce high-frequency artefacts in the discrete Fourier transform. Figure~\ref{fig:mm} summarizes the resulting mismatch distributions and separates physical modeling differences from numerical and implementation effects, to contextualize our results.

The largest mismatches arise from \texttt{TEOBResumS} vs.\ \texttt{SEOBNRv5T}, with a median of a few $10^{-3}$ and a tail above $10^{-2}$. Our results are consistent with previous studies~\cite{Abac:NRT,Haberland:2025luz} which show that mismatches between BNS models can reach the $10^{-2}$ level or higher, particularly for spinning systems, due to differences in the spinning matter and point-particle description, as well as the underlying waveform modeling assumptions. Note that we compare only to \texttt{TEOBResumS-Dal\'i}, instead of an \texttt{NRTidalv3} model or \vfourt, because \texttt{TEOBResumS} exhibits the smallest mismatches with \vfivethm\ across the same parameter space and among all BNS waveform approximants studied in Ref.~\cite{Haberland:2025luz}.

The comparison of including or omitting the $(2,1)$ and $(3,3)$-modes in \texttt{SEOBNRv5THM} yields a lower median of a few $10^{-4}$ with a tail approaching $10^{-2}$. The tail toward zero corresponds to equal-mass systems, where these odd-$m$ modes vanish. Note that these mismatches correspond to the situation where one tries to recover a signal containing the (2,1) and (3,3)-mode with a (2,2)-mode-only waveform approximant. We therefore find that neglecting higher-modes can introduce biases already at the SNR of LVK detections based on the indistinguishability condition of Eq.~\eqref{eq:indistinguishability}. We therefore strongly recommend including at least the (2,1) and (3,3)-mode for accurate PE of BNS signals in current and planned detectors.

Including the additional $(3,2)$, $(4,4)$, $(4,3)$, and $(5,5)$ modes on the other hand reduces typical mismatches to $10^{-5}$, indicating that once the leading higher-order modes are included, the remaining harmonics are quantitatively subdominant for BNS systems and do not affect PE. This implies that only the first few subdominant higher-order modes need to be included for accurate PE of BNS signals at high SNR. Given that the even-$m$ (4,4) and (3,2)-mode is non-zero for equal-mass systems, with the (4,4)-mode being dominant, it is most likely responsible for the higher mismatches at the $10^{-4}$ level.

We additionally investigate the effect of sampling rate, as most LVK BNS analyses are truncated at $f_{\rm max}\simeq 2$~kHz and typically performed with a sampling frequency of 4~kHz to reduce computational cost. Since the detectors record data at 16~kHz during normal operations, this usually entails a down-sampling of both the strain data and the waveform model. This down-sampling is known to introduce aliasing if significant power is present above the Nyquist frequency (2 kHz for a 4 kHz sampling frequency), leading to spurious contamination of the frequency band relevant for PE in BNS. While this procedure is standard in BNS analyses, its quantitative impact on mismatches has not been systematically assessed in the context of state-of-the-art EOB models.

When we compare \texttt{SEOBNRv5T} sampled at 4 kHz to it being sampled at 16 kHz, we find mismatches at the $\sim10^{-6}$ level, and conclude that sampling at 4 kHz does not introduce significant aliasing for most BNS configurations. This indicates that the waveform power above 2 kHz is sufficiently small due to its short-lived nature for typical BNS mergers, and is further down-weighted by the detector noise spectrum, such that any residual aliasing effects remain well below the level relevant for PE at realistic SNR. We do however observe a long tail of mismatches up to the $10^{-3}$ level, which corresponds to low-mass, highly spinning systems with anti-aligned spins $\chi\lesssim -0.25$ where the merger occurs at very high frequencies, leading to more power above 2 kHz and therefore more significant aliasing effects when down-sampling to 4 kHz. For these systems, we therefore recommend sampling at 16 kHz to avoid potential biases in PE.

In summary, our frequency-domain approach introduces errors that are well below the modeling uncertainties, which dominate the waveform disagreement across the explored parameter space. The speed-up we have achieved in the previous section can therefore immediately be used in PE, which we perform in the next section. 
The impact of higher modes is subleading but non-negligible, in particular for the first subdominant modes (2,1), (3,3), and perhaps (4,4) if one studies signals at high SNR $\gtrsim 60$. The frequency-domain reformulation and sampling choices introduce errors at or below the level of typical numerical effects. Any potential PE bias at realistic SNRs is therefore expected to be driven primarily by differences between waveform models rather than by our frequency-domain implementation.

\section{Fast parameter estimation} \label{sec:pe}

As a final validation of the applicability of the \vfivethm\ model for GW data analysis, we perform full Bayesian PE studies for the binary neutron star event GW170817 and for two GW170817-like injections of synthetic signals in an O5-like detector configuration. In addition to the dominant $(2,2)$ mode, we also consider analyses including the higher-order modes $(3,3)$ and $(2,1)$ in order to assess the computational performance and robustness of the model in more demanding settings.

PE is carried out using \texttt{serial Bilby} (\texttt{sBilby})~\cite{Ashton:2018jfp} and \texttt{parallel Bilby} (\texttt{pBilby})~\cite{Smith:2019ucc}, both employing the nested sampler \texttt{dynesty}~\cite{Speagle:2019ivv}. These codes serve to obtain the posterior distribution $p(\bm{\theta}\,|\,d)$ over the source parameters $\bm{\theta}$ through the application of Bayes' theorem,
\begin{equation}
    p(\bm{\theta}\,|\, d) = 
    \frac{p(d\,|\,\bm{\theta})\,p(\bm{\theta})}{E(d)},
\end{equation}
where $p(d\,|\,\bm{\theta})$ is the Gaussian-noise likelihood constructed from the strain data and the detector power spectral density (cf.\ Sec.~\ref{subsec:likelihood}), $p(\bm{\theta})$ the prior distribution, and $E(d)$ the Bayesian evidence.

For GW170817 we use the publicly available open data~\cite{LIGOScientific:2019lzm} and adopt the low-spin prior $|\chi|\leq 0.05$ employed in the LVK analyses as is done and discussed in Refs.~\cite{Haberland:2025luz,Abac:NRT,Abac:2025brd}. We follow the standard analysis settings of the LVK Collaboration~\cite{LIGOScientific:2018hze, LIGOScientific:2017vwq, Romero-Shaw:2020owr, Dietrich:2020efo, LIGOScientific:2020aai, Ashton:2021cub}, including the frequency range $20$ to $2048\,\mathrm{Hz}$, with corresponding sampling frequency $f_s=4096$ Hz, and priors uniform in detector-frame component masses $m_{1,2}\in[0.5,4]\,M_\odot$, restricted to chirp-masses in $\mathcal{M}_c\in[1.184, 1.22]$ and mass-ratios $q\in[0.05,1]$. We further use aligned-spin priors with uniform magnitudes $\chi_{1,2}\in[0,0.05]$ and isotropic orientations projected onto the orbital angular momentum vector, and uniform tidal deformabilities $\Lambda_{1,2}\in[0,5000]$. We sample luminosity distance uniform in comoving volume over $[1,75]$~Mpc, and sample over an isotropic sky location and angular-momentum orientation. Note that we recover the sky-position of GW170817 from the data rather than fixing it to the electromagnetic counterpart. For the O5-like study, we inject a low-spin, spin-aligned GW170817-like system and recover it with the same priors, except for the chirp-mass being constrained to be in $\mathcal{M}_c\in[1.15, 1.185]$ and the mass-ratio to be in $q\in[0.3,1]$.

\setlength{\extrarowheight}{8pt}
\begin{table}[h!]
\centering
\begin{ruledtabular}
\begin{tabular}{l c cc}
Parameter & Injection & \texttt{FD} & \texttt{v4T\_surr} \\
\hline
$M/M_\odot$ & 2.70 & $2.684^{+0.010}_{-0.003}$ & $2.685^{+0.014}_{-0.005}$ \\
$m_1/M_\odot$ & 1.5 & $1.41^{+0.06}_{-0.05}$ & $1.42^{+0.08}_{-0.06}$ \\
$m_2/M_\odot$ & 1.2 & $1.28^{+0.04}_{-0.05}$ & $1.27^{+0.05}_{-0.06}$ \\
$1/q$ & 1.250 & $1.10^{+0.10}_{-0.07}$ & $1.12^{+0.12}_{-0.08}$ \\
$\Lambda_1$ & 420 & $400^{+300}_{-300}$ & $400^{+270}_{-270}$ \\
$\Lambda_2$ & 730 & $700^{+400}_{-500}$ & $600^{+400}_{-400}$ \\
$\chi_{\mathrm{eff}}$ & 0.016 & $0.009^{+0.004}_{-0.002}$ & $0.010^{+0.006}_{-0.003}$ \\
$\chi_{1z}$ & 0.02 & $0.010^{+0.013}_{-0.011}$ & $0.012^{+0.013}_{-0.012}$ \\
$\chi_{2z}$ & 0.01 & $0.008^{+0.014}_{-0.012}$ & $0.009^{+0.014}_{-0.012}$ \\
$\iota/\mathrm{rad}$ & 1.47 & $1.47^{+0.04}_{-0.04}$ & $1.47^{+0.04}_{-0.04}$ \\
$d_L/\mathrm{Mpc}$ & 40 & $40.7^{+2.1}_{-2.0}$ & $40.7^{+2.1}_{-2.0}$ \\
$\phi_{\mathrm{ref}}/\mathrm{rad}$ & 1.3 & $4^{+2}_{-3}$ & $2.0^{+2.9}_{-1.7}$ \\
$\psi/\mathrm{rad}$ & 1.0 & $1.1^{+1.5}_{-0.1}$ & $1.0^{+1.5}_{-0.1}$ \\
$\alpha/\mathrm{rad}$ & 1.7 & $1.700^{+0.012}_{-0.011}$ & $1.700^{+0.012}_{-0.012}$ \\
$\delta/\mathrm{rad}$ & 0.2 & $0.199^{+0.011}_{-0.013}$ & $0.199^{+0.012}_{-0.013}$ \\
$\rho^{\mathrm{H1}}_{\mathrm{mf}}$ & 33.341 & $33.22^{+0.07}_{-0.17}$ & $33.23^{+0.07}_{-0.18}$ \\
$\rho^{\mathrm{L1}}_{\mathrm{mf}}$ & 24.134 & $24.03^{+0.06}_{-0.14}$ & $24.03^{+0.06}_{-0.14}$ \\
$\rho^{\mathrm{V1}}_{\mathrm{mf}}$ & 33.705 & $33.60^{+0.06}_{-0.17}$ & $33.60^{+0.07}_{-0.17}$ \\
$\log \mathcal{BF}$ & -- & $1364.50 \pm 0.23$ & $1364.42 \pm 0.23$ \\
\end{tabular}
\end{ruledtabular}
\caption{Injected and recovered values for the low aligned-spin BNS O5 injection and recovery with \vfivethmfd\ and \texttt{SEOBNRv4T\_surrogate}. Uncertainties stated are based on the 16\%, 50\% and 84\% quantiles. We only state the recovered values of the full \texttt{FD} recovery, as the posteriors resulting from the other methods were statistically indistinguishable. We include the matched-filter SNRs $\rho_{\rm mf}$ for the individual detectors as well as the Bayes factor $\mathcal{B}\mathcal{F}$ of the signal hypothesis.}
\label{tab:injection_settings}
\end{table}

\setlength{\extrarowheight}{2pt}
\begin{table*}
\centering
\setlength{\tabcolsep}{9pt}
\begin{tabular}{
l
c
c
l
l
l
c}
\hline
\hline
System & Mode content &  SNR  & Sampler & \makecell[ll]{Changes to \\ model/likelihood} & Runtime & Efficiency \\
\hline

\multirow{6}{*}{GW170817} & \multirow{6}{*}{$(2,2)$} & \multirow{6}{*}{33.3} 
& \multirow{4}{*}{\texttt{sBilby}} 
& --- & \runtime{4}{0}{7} & 1.0 \\
& & & & \FD  & \runtime{1}{20}{43} & 2.1 \\
& & & & \FD\ Multibanding  & \runtime{1}{0}{20} & 3.9 \\
& & & & \FD\ Relative binning & \runtime{1}{0}{29} & 3.9 \\
\cline{4-7}
& & & \multirow{2}{*}{\texttt{pBilby}}
& --- & \runtime{1}{5}{25}& 0.4 \\
& & & & \FD & \runtime{0}{21}{48}& 0.5 \\
\hline

\multirow{4}{*}{GW170817} & \multirow{4}{*}{$(2,2),\,(3,3),\, (2,1)$} & \multirow{4}{*}{33.4}
& \multirow{4}{*}{\texttt{sBilby}}
& --- & \runtime{6}{1}{57} & 1.0 \\
& & & & \FD  & \runtime{2}{6}{55} & 2.7 \\
& & & & \FD\ Multibanding & \runtime{1}{2}{40} & 5.5 \\
& & & & \FD\ Relative binning & \runtime{1}{2}{41} & 5.5 \\
\hline

\multirow{6}{*}{BNS injection} 
& \multirow{6}{*}{$(2,2),\,(3,3),\, (2,1)$} & \multirow{6}{*}{53.1}
& \multirow{6}{*}{\texttt{sBilby}}
& --- & \runtime{12}{0}{9} & 1.0 \\
& & & & \FD  & \runtime{6}{12}{40} & 1.8 \\
& & & & \FD\ Multibanding  & \runtime{1}{16}{30} & 7.1 \\
& & & & \FD\ Relative binning & \runtime{1}{16}{29} & 7.1 \\
\cline{5-7}
& & & & \texttt{v4T\_surr}  & \runtime{4}{16}{0} & 2.6 \\
& & & & \texttt{v4T\_surr} Multibanding  & \runtime{0}{3}{16} & 26.0 \\
\hline
\hline
\end{tabular}
\caption{
Comparison of PE runtimes for different model and likelihood implementations, where we indicate our frequency-domain model as \texttt{FD} to differentiate it from the conventional time-domain implementation. We re-analyze GW170817 with and without higher modes, and perform a PE analysis of a synthetic BNS signal injection with O5 design sensitivity. For the latter, we also perform a recovery with \texttt{SEOBNRv4T\_surrogate} (abbreviated as \texttt{v4T\_surr}). \texttt{IMRPhenomXAS\_NRTidalv3} takes nearly the same time as the surrogate, and is not included in the table. Reported times correspond to walltime. Efficiency is defined as relative CPU time (walltime$\,\times\,$number of cores) compared to the FFT likelihood for the same system and mode content.
}
\label{tab:pe_runtime_comparison}
\end{table*}

All analyses are performed with $1000$ live points, number of accepted jumps per MCMC chain naccept$\,=60$ and the \texttt{acceptance walk} method of \texttt{dynesty}. When using \texttt{sBilby} we employ $64$ cores, while \texttt{pBilby} runs using $16$ nodes with $32$ cores each, corresponding to eight times more computational resources.

We study three setups in which we perform PE. For GW170817, we perform PE once when just using the (2,2)-mode version of our waveform model, \texttt{SEOBNRv5T} as was done previously in Ref.~\cite{Haberland:2025luz}, and once when including the higher-order modes (2,1) and (3,3) with \texttt{SEOBNRv5THM}. For the O5 synthetic signals, we only perform PE with both injection and recovery containing the (2,2), (2,1) and (3,3)-modes. We decide on only including these three modes as a test of our higher-mode implementation and justify it through the results of Sec.~\ref{subsec:mismatches}, where we find that neglecting the (3,2), (4,4), (4,3) and (5,5) modes can introduce mismatches at and below the $10^{-4}$ level, which should not affect the results of this section due to the SNRs that we consider. Given that the systems we are studying have mass ratios $q\lesssim 2$, one could include in particular the (4,4) and (3,2)-modes in real PE analyses in the future. In comparison to previous analyses of GW170817, we also use a lower frequency cutoff of 20 Hz instead of 23 Hz, and find a higher optimal SNR of 33.4 compared to the reported network SNR of 32~\cite{LIGOScientific:2018hze}.

The injected and recovered parameters for the synthetic signal in an O5 detector-network are stated in Table~\ref{tab:injection_settings}, which we inject into an HLV network with zero-noise, where Hanford and Livingston have as power spectral density the design target of O5 and Virgo has the high limit target sensitivity for O5, both from Ref.~\cite{LVK_psds}. The system corresponds to a GW170817-like event in O5, meaning that it has similar masses, very low, aligned spins, as well as a similar distance at 40 Mpc. The main differences between GW170817 and the O5 synthetic signals are that the latter is placed in a more sensitive detector network and includes a non-negligible effective spin, resulting in a longer signal (256~s) and higher SNR, which allows us to assess the performance of our model in a more demanding regime. It is also more edge-on to increase the effect of higher-order modes and has an arbitrarily chosen sky position. We also perform a recovery of the O5 synthetic signals with the surrogate model \texttt{SEOBNRv4T\_surrogate} and the phenomenological \texttt{IMRPhenomXAS\_NRTidalv3} (abbreviated as \texttt{PhenomXAS\_NRTidalv3} in some of the following figures) to assess the effect of waveform systematics and speed in this context.

In Ref.~\cite{Haberland:2025luz}, it was found that \texttt{pBilby} can be much less efficient compared to the serial case. For the (2,2)-mode-only re-analysis of GW170817, we intend to quantify this by comparing the efficiency of \texttt{sBilby} to a re-analysis using \texttt{pBilby}.

\subsection{Likelihood implementations and speed-ups}

To assess computational performance, we compare several likelihood implementations. Besides the standard time-domain FFT-based likelihood, we employ our frequency-domain implementations in (i) the full likelihood (i.e., we evaluate it on the same frequency-grid the FFT is also evaluated on), (ii) the multibanding likelihood, and (iii) the likelihood using relative binning. The two latter approaches are reviewed in Sec.~\ref{sec:mb_rb}, and are here applied in a full PE context. The fiducial waveform for the relative binning method has parameters $m_1=1.465\ M_{\odot}$, $m_2=1.292\ M_{\odot}$, $\chi_1=\chi_2=0$, $\Lambda_2^{(1)}=420$, $\Lambda_2^{(2)}=680$, $d_L=40$, $\iota=2.6$, $\varphi=\psi=0$ and the correct sky-location as determined by the electromagnetic counterpart for GW170817. When using the relative binning method for the synthetic signals, we recover with the fiducial waveform being evaluated at the injected values.

Since the primary goal of this study is to reduce the computational cost of PE, Table~\ref{tab:pe_runtime_comparison} summarizes the walltimes required for each configuration and the efficiency as compared to the standard FFT approach. For a fixed system and mode content, the reported efficiency is defined as the CPU time relative to the standard FFT likelihood. Several trends are apparent.

First, moving from the time-domain implementation to a frequency-domain evaluation already reduces the runtime noticeably as the SPA outperforms the standard FFT approach. Incorporating multibanding or relative binning yields a further substantial reduction. For GW170817 with only the dominant $(2,2)$ mode, we observe speed-ups of a factor $\sim 2$--$3$ compared to the baseline FFT likelihood. When including higher-order modes, the gains become more pronounced, reaching factors of around six. The same qualitative behavior is observed for the O5-like synthetic signals, where the overall runtime is larger due to the increased signal duration and detector sensitivity, but multibanding and relative binning again reduce the computational cost by close to a factor of seven.
Second, the relative performance of multibanding and relative binning is comparable across all configurations, with small differences at the $10\%$ level. Given that fewer assumptions are used in the multibanding technique, we therefore recommend its use with \texttt{SEOBNRv5THM\_FD}.

Finally, as was found in Ref.~\cite{Haberland:2025luz}, \texttt{pBilby} achieves slightly shorter walltimes due to increased computational resources, though at the cost of significantly more cores. We also find that the speed-improvement of our frequency-domain method does not affect the walltime of the \texttt{pBilby} analysis much, in agreement with the above reference. The scaling behavior is consistent with expectations from parallel nested sampling, and the efficiency with respect to CPU time is at the 50\% level compared to the \texttt{sBilby} analyses, most likely due to communication overhead and parallel nested sampling scaling, as well as the fact that the 32-core CPUs we have used (AMD EPYC 7351) are also slightly less efficient than our 64-core CPUs (AMD EPYC 7513). The total walltime of \texttt{pBilby} is furthermore nearly identical to the walltime of \texttt{sBilby} when using multibanding or relative binning, showing that either method can be used to perform PE in a reasonable timescale. Due to its computational efficiency, we would, however, suggest using \texttt{sBilby} for BNS analyses employing \texttt{SEOBNRv5THM}.

\begin{figure}[t]
    \centering
   \includegraphics[width=0.99\linewidth]{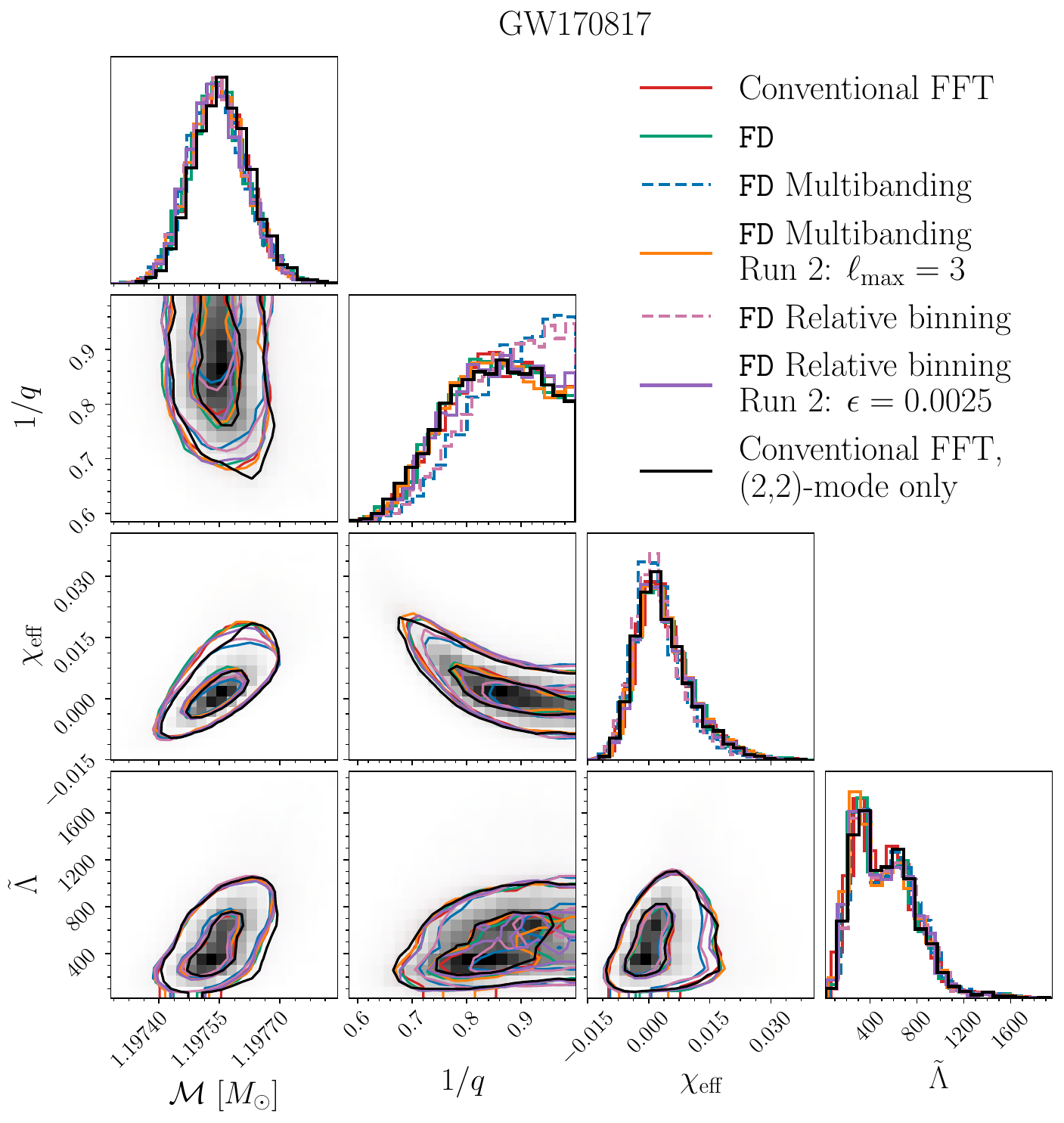}
   \caption{The resulting posteriors of our re-analysis of GW170817 with and without higher-modes being considered. For visibility reasons we include only the FFT posterior of the (2,2)-mode-only run, as all (2,2)-mode-only posteriors were statistically indistinguishable. The two runs where the mass-ratio has a different posterior (dashed) are discussed in the main text.}
   \label{fig:GW170817}
\end{figure}

\begin{figure}[t]
    \centering
   \includegraphics[width=0.99\linewidth]{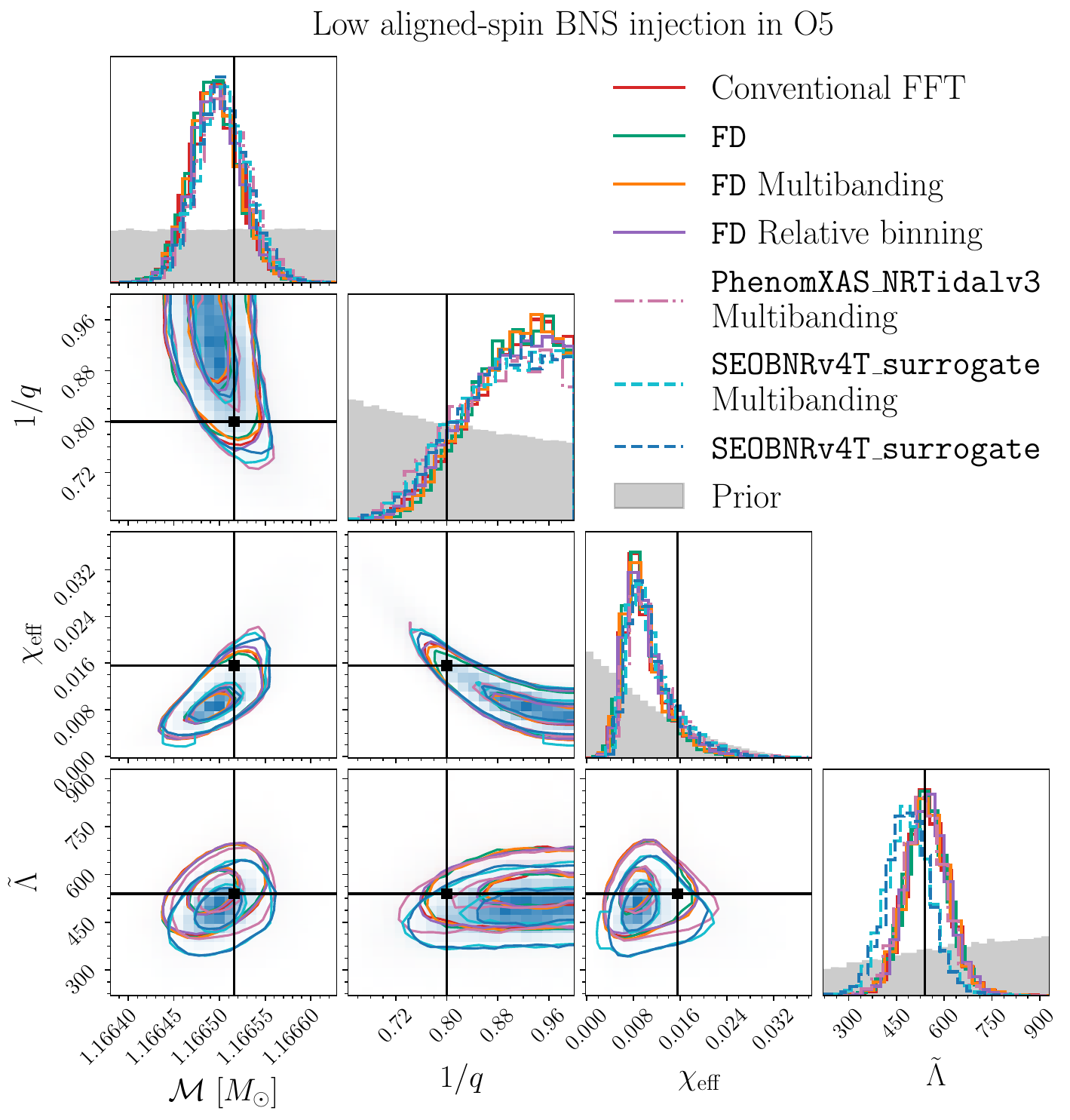}
   \caption{The resulting posteriors, and truth values of our synthetic signal of a low-spin, spin-aligned GW170817-like signal in an O5 HLV detector network. We always inject the time-domain waveform model \texttt{SEOBNRv5THM} including higher-modes and recover with \texttt{SEOBNRv5THM}, as well as, the (2,2)-mode only approximants \texttt{SEOBNRv4T\_surrogate} and \texttt{IMRPhenomXAS\_NRTidalv3}. We indicate the prior for our analysis as a grey region.}
   \label{fig:Injection}
\end{figure}

\subsection{Posterior consistency and bias assessment}

The marginalized one- and two-dimensional posterior distributions for a representative subset of intrinsic parameters are shown in Figs.~\ref{fig:GW170817} and \ref{fig:Injection}. In both cases, we display the detector-frame chirp mass $\mathcal{M}_c$, the inverse mass-ratio $1/q$, the effective spin $\chi_{\rm eff}$, and the combined tidal deformability $\tilde{\Lambda}$.

Across all likelihood implementations --- time domain, frequency domain, multibanded, and relative binning --- we find statistically indistinguishable posterior distributions if applied correctly, which is also the case between (2,2)-mode-only and higher-mode content analyses. The medians and credible intervals agree within sampling uncertainty, and no systematic shifts are observed in any parameter. In particular, the tight constraints on the chirp mass and the broader posteriors in tidal and spin parameters are consistently reproduced.

For both multibanding and relative binning we initially encountered issues when analyzing GW170817 with higher-mode waveform models due to different reasons. To illustrate the sensitivity of these accelerated likelihood methods to higher-mode content, we show the corresponding biased posteriors as dashed lines in Fig.~\ref{fig:GW170817} and note that, while all methods agree if applied correctly, we intend to highlight the subtle pitfalls we encountered. For the multibanding method, while we have included higher order modes in both our synthetic signal and the recovery, we did not include this information in the likelihood by setting $\texttt{highest\_mode}=3$ in the likelihood arguments. For the relative binning method on the other hand, we did not include sufficient frequency points, as the inclusion of higher-modes introduced extra structure into the waveform modes, which breaks the assumption of Eq.~\eqref{eq:linear_ratio}. One can however mitigate this issue by reducing the $\epsilon$ parameter by a factor of 10 compared to its default value (i.e., by setting $\epsilon=0.0025$). Note that there also exists a version of multibanding that is performed on a mode-by-mode basis and is therefore less affected by the inclusion of higher-order modes~\cite{Leslie:2021ssu}, which we however did not use.

For the study of synthetic signals, we only run the corrected methods and find that the injected parameters are recovered well for all likelihood implementations as can be seen from Table~\ref{tab:injection_settings}, where we include the recovered values of the full frequency-domain likelihood and the 1-$\sigma$ confidence intervals.

We observe a mild systematic shift in the recovered mass ratio $q$ and effective spin, consistent with previous studies (e.g., Ref.~\cite{Abac:2025brd}). This behavior can be understood from the well-known degeneracy between $q$ and $\chi_{\rm eff}$ entering the phase evolution at 1.5PN order which are well constrained due to the length of the signal. Because the prior on $\chi_{\rm eff}$ is furthermore strongly peaked at zero (via the low-spin prior), whereas the prior on $q$ is approximately uniform over the posterior support, the inference is biased toward smaller $\chi_{\rm eff}$, as can be seen from the included prior in Fig.~\ref{fig:Injection}. Through the degeneracy, a smaller $\chi_{\rm eff}$ is then countered by a bias towards equal mass, which shifts probability density towards more equal-mass configurations. The observed posterior displacement is therefore consistent with prior-volume effects rather than a modeling bias. It is furthermore very interesting to see how well the spins can be measured for the individual NSs, which is a direct consequence of the long signal duration and the high SNR in O5, despite the fact that they are at a very low $\chi_{1z}=0.02$ and $\chi_{2z}=0.01$, respectively.

We also find a slightly biased recovery of the tidal deformabilities only when recovering with \texttt{SEOBNRv4T\_surrogate}, but not when recovering with \texttt{IMRPhenomXAS\_NRTidalv3} or \vfivethm. This is expected due to the high mismatches found also in Refs.~\cite{Haberland:2025luz,Abac:NRT} and further in agreement with the systematic studies of Refs.~\cite{Gamba:2021prd,Kunert:2024}. The bias can be attributed to the lower amount of analytical information in the tidal and point-particle sector of \texttt{SEOBNRv4T}, different modeling choices and different employed quasi-universal relations for the free NS parameters. This should also serve as a testament to the importance of using accurate waveform models for PE, as the systematic bias in the tidal deformabilities is significant compared to the statistical uncertainty, and could lead to incorrect conclusions about the NS EoS if not properly accounted for. To this end, we list in the first row of Table~\ref{tab:mismatches_injection} the mismatches of Eq.~\eqref{eq:mismatch} of the two approximants against \vfivethm\ including higher-order modes evaluated at the injected parameters, as well as the corresponding SNRs at which one would expect to see a bias in PE based on Eq.~\eqref{eq:indistinguishability}, which is consistent with the observed bias in the posteriors in Fig.~\ref{fig:Injection}.

\begin{table}[t]
\centering
\begin{ruledtabular}
\begin{tabular}{l l c c}
Injection & Waveform Model & $1-\mathcal{M}$ & $\rho_{1\sigma}$  \\
\hline
\multirow{2}{*}{Low aligned} & \texttt{PhenomXAS\_NRTidalv3} & $3.980\times10^{-4}$ & 86.8  \\
& \texttt{SEOBNRv4T\_surrogate} & $1.089\times10^{-3}$ & 52.5  \\
\hline
\multirow{2}{*}{Med. anti-aligned} & \texttt{PhenomXAS\_NRTidalv3} & $5.908\times10^{-4}$ & 71.3  \\
& \texttt{SEOBNRv4T\_surrogate} & $6.234\times10^{-4}$ & 69.4 \\
\end{tabular}
\end{ruledtabular}
\caption{Mismatches and corresponding SNRs at which one would expect to see a bias in PE for the two approximants used for the recovery of the O5 synthetic signal. We compare the mismatches of the approximant against \vfivethm\ including higher-order modes evaluated at the injected parameters.}
\label{tab:mismatches_injection}
\end{table}

\begin{figure}[t]
    \centering
   \includegraphics[width=0.99\linewidth]{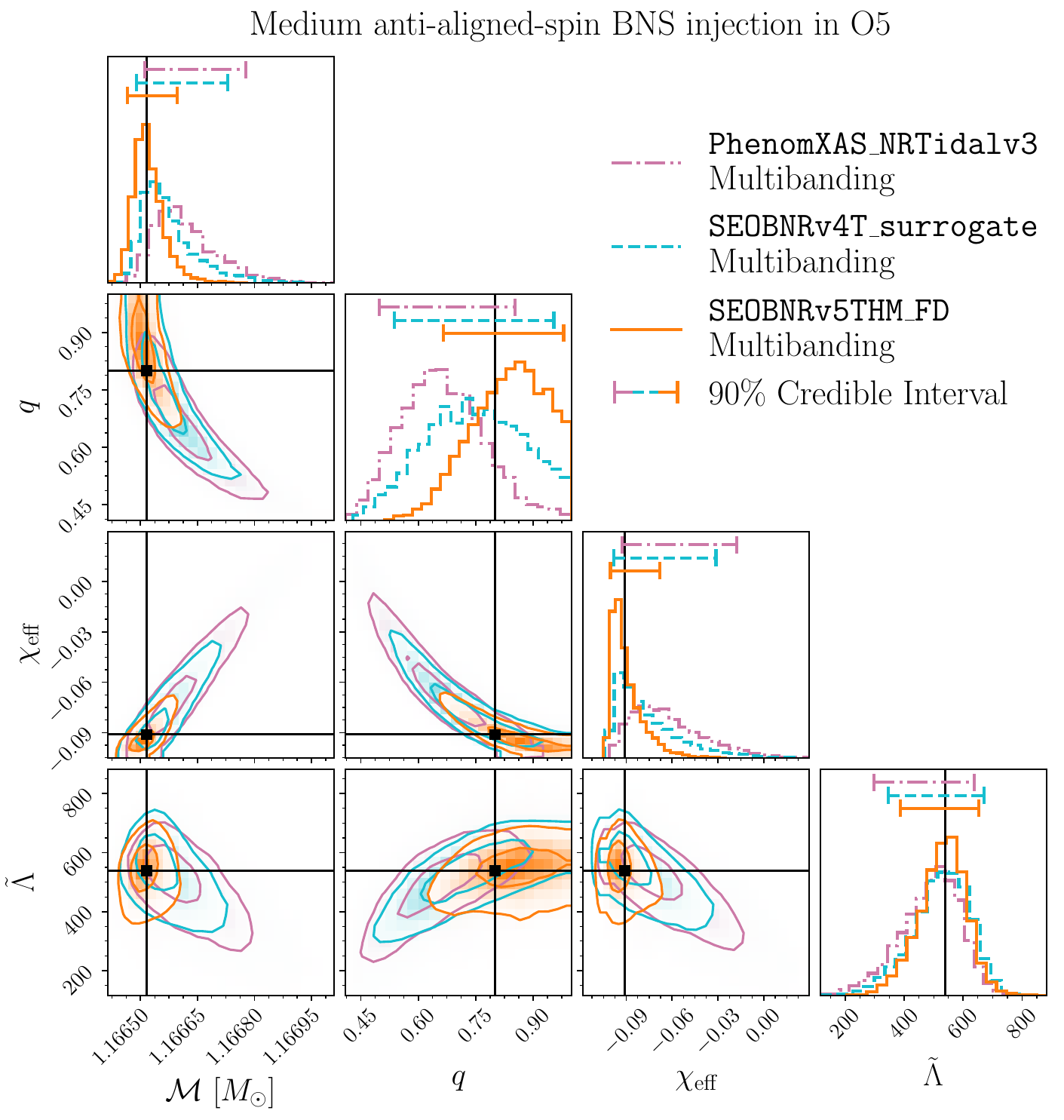}
   \caption{The resulting posteriors, and truth values of our analysis of the synthetic signal of an anti-aligned GW170817-like signal in an O5 HLV detector network. We always inject the time-domain waveform model \texttt{SEOBNRv5THM} with higher-modes and recover with it, as well as the (2,2)-mode only approximants \texttt{SEOBNRv4T\_surrogate} and \texttt{IMRPhenomXAS\_NRTidalv3}. We use multibanding in all cases and indicate the 90\% credible intervals for the recovered parameters.}
   \label{fig:Antialigned_Injection}
\end{figure}

As a final part of our study, we also inject an even more demanding anti-aligned system with significant anti-aligned spins of $\chi_{1z}=-0.1$ and $\chi_{2z}=-0.08$ in the same O5 HLV network. We keep all other parameters the same as the previous injection, and recover again with the (2,2)-mode only approximants \texttt{SEOBNRv4T\_surrogate} and \texttt{IMRPhenomXAS\_NRTidalv3}, as well as \vfivethmfd, where we include higher-order modes. We find that one needs to increase the prior in spin magnitude to $\chi_{1,2}\in [0, 0.4]$ in order to correctly recover the injected value for all waveform models, and use multibanding for the recovery in all cases. The resulting posteriors are shown in Fig.~\ref{fig:Antialigned_Injection}. 

We find that the bias in the mass ratio and therefore individual masses and the effective tidal deformability is even more pronounced than for the previous synthetic signal, which is expected as \vfivethm\ includes higher-order modes and spin-shifted dynamical tides, which are known to more strongly affect anti-aligned systems~\cite{Steinhoff:2021dsn,Haberland:2025luz} and break the degeneracy between $q$ and $\chi_{\rm eff}$. In contrast to the analysis of the synthetic signal of Fig.~\ref{fig:Injection}, where \texttt{IMRPhenomXAS\_NRTidalv3} recovered unbiased posteriors, we instead find biased posteriors in all relevant parameters for both \texttt{IMRPhenomXAS\_NRTidalv3} and \texttt{SEOBNRv4T\_surrogate} in this case. This is noteworthy, as the mismatches in the lower row of Table~\ref{tab:mismatches_injection}, are very low and do not suggest the strong biases we observe. As both masses and tidal deformabilities are used to draw conclusions about the NS EoS, this again highlights the importance of using different waveform models for PE, to understand the effects of waveform systematics in future events and to draw correct conclusions about spinning BNS systems if detected in the future. It further highlights shortcomings of the commonly used criterion of Eq.~\eqref{eq:indistinguishability} to assess waveform systematics, as it does not take into account the effect of parameter degeneracies, or the prior volume, in particular for long signals.

Taken together, these results demonstrate that the frequency-domain implementations, including multibanding and relative binning, yield posterior distributions fully consistent with the standard time-domain analysis while significantly reducing computational cost. The speed-ups become increasingly important when including higher-order modes or analyzing longer, louder signals expected in future observing runs. We also find that one cannot rely only on waveform approximants used for previous runs, as BNS waveform systematics can already lead to non-negligible biases in current-generation detectors for individual events, and will become even more important in the future. The ability to perform PE with the full \vfivethm\ model in a reasonable timescale is therefore crucial for robust inference of NS properties and the EoS.
Going forward, our frequency-domain implementations will furthermore make it feasible to study signals in ET/CE detectors of multiple hours when used in tandem with the multibanding or relative binning method, which would be fully impossible with standard techniques as one can see from Fig.~\ref{fig:benchmark}. To this end, one would need to take Earth's rotation into account during PE however.

\section{Conclusion} \label{sec:conclusion}

In this work, we have developed a frequency-domain implementation of the \vfivethm\ waveform model for quasi-circular, spin-aligned BNS systems~\cite{Haberland:2025luz}. It is designed to meet the dual requirements of high accuracy and computational efficiency for modern and future GW data analysis. Our approach combines the SPA for the early inspiral with an FFT treatment of the late inspiral and post-inspiral regime, applied consistently on a mode-by-mode basis. This hybrid construction leverages the strengths of both techniques while mitigating their individual limitations, yielding a flexible and robust frequency-domain waveform model that consistently incorporates higher-order modes.

We have demonstrated that this method accurately reproduces the underlying time-domain \vfivethm\ waveforms, independent of the higher-order mode content. In particular, our transition scheme between SPA and FFT, based on physically motivated criteria, ensures that the loss of accuracy inherent to the SPA near merger is avoided. This is especially important as matter effects are expected to become dominant in the late stages of the binary coalescence.

Beyond computational performance, our implementation preserves the full physical content of the \vfivethm\ model, including state-of-the-art treatments of tidal effects, spin-induced multipole moments, dynamical tides, and higher harmonics, as well as calibration to numerical relativity. As such, it provides a faithful and efficient tool for extracting neutron-star properties, including tidal deformabilities, from GW observations, thereby contributing directly to efforts to constrain the equation of state of dense nuclear matter.

The resulting frequency-domain waveforms show excellent agreement with full FFT-based transforms, with mismatches remaining well below the level required for PE in current and next-generation detectors across the relevant BNS parameter space. This accuracy is further validated through parameter estimation studies of GW170817 and a GW170817-like synthetic signal in an O5 detector network, where we find that the posteriors obtained with our frequency-domain implementation are statistically indistinguishable from those obtained with the standard time-domain FFT-based likelihood, while achieving significant reductions in computational cost. Due to the generality of this approach, it is furthermore applicable to BBH systems as well, with potential implications for data analysis challenges in ET/CE as well as LISA.

By enabling direct evaluation on non-uniform frequency grids, our frequency-domain construction allows for seamless integration with accelerated likelihood techniques, such as multibanding~\cite{Morisaki:2021ngj,Cornish:2021lje} and relative binning~\cite{Zackay:2018qdy,Krishna:2023bug}. When combined with these methods, we achieve reductions in the overall computational cost of PE for signals in LVK detectors at close to an order of magnitude, with even larger gains expected for the long-duration signals of ET/CE detectors. PE with the full \vfivethm\ model is therefore enabled in the timescale of a few days.

A central result of this work is the identification and mitigation of the dominant computational bottleneck in waveform generation, namely the waveform interpolation cost, with the new bottleneck becoming the EOB dynamics integration, which currently requires $\mathcal{O}(0.1\,\mathrm{s})$ per waveform. This currently prohibits us from reaching the computational efficiency of phenomenological or surrogate models, which can be evaluated in milliseconds, with full PE being performable in hours. We do however achieve the ability to run PE with the full \vfivethm\ model in the timescale of a few days.

Our results further highlight two important physical implications for GW inference. First, we find that higher-order modes can have a measurable impact on PE, challenging the common assumption that they can be safely neglected in binary neutron star analyses. Second, our PE analyses of synthetic signals demonstrates that waveform systematics, as illustrated by comparisons with the \texttt{SEOBNRv4T\_surrogate}~\cite{Lackey:2018zvw} and \texttt{IMRPhenomXAS\_NRTidalv3}~\cite{Abac:NRT}, can already lead to non-negligible biases in current-generation detectors for individual events. These findings emphasize that improvements in waveform modeling are not only relevant for future observatories, but are already important for robust inference with data from current GW  observation runs. This furthermore highlights the importance of being able to obtain results from different waveform approximants in $\mathcal{O}$(days).

Looking ahead, the framework introduced here opens several avenues for further development. Extensions to more general configurations, including precession and eccentricity, as well as refinements of the SPA to the next order or with the shifted uniform asymptotics, could further enhance both accuracy and efficiency. The extension of our approach to the massive BH binary case for LISA data-analysis would be worthwhile to investigate, in particular in the context of time-frequency analyses~\cite{Cornish:2020odn}. In addition, the compatibility of our approach with emerging inference techniques, including hardware-acceleration and machine-learning-based methods, positions it as a versatile component in the evolving landscape of GW data analysis.

In summary, our work provides a crucial step toward bridging the gap between highly accurate time-domain EOB models and the computational demands of large-scale PE. By enabling fast, flexible, and precise waveform generation in the frequency domain via \vfivethmfd, we equip the community with a tool that is well suited for the challenges posed by the rapidly increasing sensitivity and data volume of current and future GW observatories, while simultaneously improving the robustness of astrophysical inference against waveform systematics.

\section*{\label{section: Acknowledgments}Acknowledgments}

We thank Gonzalo Morras and Koustav Chandra for
sharing their general knowledge on waveforms and \texttt{Bilby}, 
and for many helpful discussions. We are grateful to  Aurora
Abbondanza for insightful discussions, in particular on long-duration
PE, and higher-order mode mismatches. We also thank Lorenzo
Pompili for his suggestions on the time-frequency representation that
have inspired this project, as well as his careful review of this manuscript. 
We thank Felix Lichtner and Trevor Scheopner
for their rigorous derivation of the SPA up to second order, as well
as insights into its systematic series expansion and relation to
Stirling's approximation.

Furthermore, we thank 
Adrian Abac,
Tim Dietrich,
Raffi Enficiaud,
H\'ector Estelles,
Guglielmo Faggioli,
Cheng Foo,
Aldo Gamboa,
Nihar Gupte,
Saketh Maddu,
Peter James Nee,
Nami Nishimura,
Sergei Ossokine,
Luca Sebastiani,
Jan Steinhoff, and
Maarten van de Meent
for helpful discussions, constructive criticism and suggestions that have improved this work.

The computational work for this manuscript was carried out on the compute cluster Hypatia at the Max Planck Institute for Gravitational Physics in Potsdam. This research has made use of data or software obtained from the Gravitational Wave Open Science Center (\href{https://gwosc.org}{gwosc.org}), a service of the LIGO Scientific Collaboration, the Virgo Collaboration, and KAGRA. This material is based upon work supported by NSF's LIGO Laboratory which is a major facility fully funded by the National Science Foundation, as well as the Science and Technology Facilities Council (STFC) of the United Kingdom, the Max-Planck-Society (MPS), and the State of Niedersachsen/Germany for support of the construction of Advanced LIGO and construction and operation of the GEO600 detector. Additional support for Advanced LIGO was provided by the Australian Research Council. Virgo is funded, through the European Gravitational Observatory (EGO), by the French Centre National de Recherche Scientifique (CNRS), the Italian Istituto Nazionale di Fisica Nucleare (INFN) and the Dutch Nikhef, with contributions by institutions from Belgium, Germany, Greece, Hungary, Ireland, Japan, Monaco, Poland, Portugal, Spain. KAGRA is supported by Ministry of Education, Culture, Sports, Science and Technology (MEXT), Japan Society for the Promotion of Science (JSPS) in Japan; National Research Foundation (NRF) and Ministry of Science and ICT (MSIT) in Korea; Academia Sinica (AS) and National Science and Technology Council (NSTC) in Taiwan.

\appendix

\section{Fourier domain conditioning of the modes} \label{app:conditioning}

To construct a frequency-domain conditioning analogous to the time-domain tapering used in the \vfive\ models, we approximate the early inspiral locally around the start time $t_0$ and derive the corresponding SPA expression for a short tapered segment.

For a given $(\ell,m)$ mode we will use the phase, its derivatives and its amplitude at the start-time, which we refer to as $\phi_0=\phi(t_0)$. We further assume $\dot f_{\ell m}^{\rm start} > 0$, as required for the validity of the SPA.

In a short interval around $t_0$ we expand the phase to quadratic order,
\begin{equation}
\Phi(t)
\simeq
\Phi_0
+ 2\pi f_{\ell m}^{\rm start} (t - t_0)
+ \pi \dot f_{\ell m}^{\rm start} (t - t_0)^2,
\end{equation}
where $\Phi_0 = \Phi_{\ell m}(t_0)$.

The stationary phase condition
\begin{equation}
2\pi f = \dot{\Phi}_{\ell m}(t)
\end{equation}
then yields the linearized frequency-time relation
\begin{equation}
\Delta t = t(f) - t_{\rm abs} = \frac{f - f_{\ell m}^{\rm start}}
{\dot f_{\ell m}^{\rm start}}.
\end{equation}

We introduce a time window of duration $T_{\rm win}$ and define the lowest frequency to be populated by evolving the linearized frequency relation backwards in time,
\begin{equation}
f_{\ell m}^{\rm cond}
=
f_{\ell m}^{\rm start}
-
\dot f_{\ell m}^{\rm start} T_{\rm win}.
\end{equation}
Below $f_{\ell m}^{\rm cond}$ the mode is set to zero.

Over the interval $t \in [t_{\rm abs}-T_{\rm win},t_{\rm abs}]$ we model the waveform as a Hanning-tapered oscillation with frozen amplitude,
\begin{equation}
h_{\ell m}^{\rm taper}(t)
=
\frac{A_0}{2}
\left[
1 - \cos \left(
\pi \frac{t - t_0}{T_{\rm win}}
\right)
\right]
e^{i \Phi_{\ell m}(t)},
\end{equation}
where $A_0 = A(t_{\rm abs})$.

Applying the SPA, the Fourier transform can be written as
\begin{equation}
\tilde h_{\ell m}(f)
\simeq
\frac{A(t_f)}{\sqrt{\dot f(t_f)}}
\exp \big[i \Psi(f)\big],
\end{equation}
with
\begin{equation}
\Psi(f)
=
\frac{\pi}{4}
+
\Phi_{\ell m}(t_f)
-
2\pi f \big(t_f - t_{\rm attach}\big),
\end{equation}
where $t_f$ satisfies the stationary phase condition.

Using the quadratic expansion, the phase becomes
\begin{align}
        \Psi(f) = & \frac{\pi}{4} + \Phi_0 + 2\pi \left[ f_{\ell m}^{\rm start} \Delta t + \frac{1}{2} \dot f_{\ell m}^{\rm start} \Delta t^2 \right] \\ & - 2\pi f \big(\Delta t + t_0 - t_{\rm attach}\big). \notag
\end{align}

Freezing the amplitude at $A_0$, the SPA amplitude in the conditioning region becomes
\begin{equation}
\tilde A_{\ell m}^{\rm cond}(f)
=
\frac{A_0}{2 \sqrt{\dot f_{\ell m}^{\rm start}}}
\left[
1
+
\cos \left(
\pi \frac{\Delta t}{T_{\rm win}}
\right)
\right].
\end{equation}
By construction,
\begin{equation}
\tilde A_{\ell m}^{\rm cond}(f_{\ell m}^{\rm start})
=
\frac{A_0}{\sqrt{\dot f_{\ell m}^{\rm start}}},
\qquad
\tilde A_{\ell m}^{\rm cond}(f_{\ell m}^{\rm cond})
= 0,
\end{equation}
so that the waveform smoothly vanishes at the lower edge of the conditioning interval and continuously matches the usual SPA expression at $f_{\ell m}^{\rm start}$.

The frequency-domain mode is therefore defined piecewise as
\begin{equation}
\tilde h_{\ell m}(f)
=
\begin{cases}

& 0,\quad f < f_{\ell m}^{\rm cond}, \\
& \tilde A_{\ell m}^{\rm cond}(f)
e^{i \Psi(f)}, \quad f_{\ell m}^{\rm cond}
\le f
< f_{\ell m}^{\rm start}, \\
& \tilde h_{\ell m}^{\rm SPA}(f), 
f \ge f_{\ell m}^{\rm start}.
\end{cases}
\end{equation}

This construction removes unphysical low-frequency content below the true start frequency of each $(\ell,m)$ mode while ensuring a smooth attachment to the inspiral SPA waveform and avoiding sharp spectral cutoffs that would otherwise introduce artefacts.

\section{Fourier domain polarizations} \label{app:polarizations}

Due to the Fourier transform and polarization conventions, one has to be careful in the derivation of the resulting polarization. The two polarizations can be obtained from the modes via
\begin{subequations}
    \begin{align}
    h_+(t) &= \mathrm{Re}[h(t)] = \frac{1}{2} \left[h(t) + h^*(t)\right]\ ,\\
    h_\times(t) &= -\mathrm{Im}[h(t)] = -\frac{1}{2i} \left[h(t) - h^*(t)\right]\ , \label{eq:hcross_td}
    \end{align}
\end{subequations}
where we can insert the spherical harmonic decomposition of the strain $h(t)$ from Eq.~\eqref{eq:hoft_sphericalH} to find
\begin{equation}
\begin{aligned}
    & h_+(t) = \frac{1}{2} \sum_{\substack{\ell\geq 2 \\ \ell \geq |m| > 0}} \,\left[ \,_{-2}Y_{\ell, m} h_{\ell, m}(t) + \,_{-2}Y_{\ell, m}^* h^*_{\ell, m}(t)\right] \\
    &= \frac{1}{2} \sum_{\substack{\ell\geq 2 \\ \ell \geq |m| > 0}} \left[ _{-2}Y_{\ell, m} h_{\ell, m}(t) + \,_{-2}Y_{\ell, m}^* (-1)^\ell h_{\ell, -m}(t)\right] \\
    &= \frac{1}{2} \sum_{\substack{\ell\geq 2 \\ \ell \geq |m| > 0}} \,\left[ \,_{-2}Y_{\ell, m} + (-1)^\ell \,_{-2}Y_{\ell, -m}^* \right] h_{\ell, m}(t) \ , 
\end{aligned}
\end{equation}
where we have applied the axial symmetry of the system $h_{\ell, m} = (-1)^\ell h^*_{\ell, -m}$ in the second line and rewrote the dummy index $m \mapsto -m$ in the last line. As a last step, we would like to express the polarizations solely in terms of the $m>0$ modes which we have studied in Sec.~\ref{subsec:SPA}. To this end, we will employ the symmetry property of the Fourier transform of Eq.~\eqref{eq:symmetry_property} as applied to real-valued functions to find 
\begin{equation} \label{eq:fd_polarizations_app}
    \begin{aligned}
    & \tilde{h}_+(f) = \tilde{h}^*_+(-f) \\
    & = \frac{1}{2} \sum_{\substack{\ell\geq 2 \\ \ell \geq |m| > 0}} \,\left[ \,_{-2}Y_{\ell, m} + (-1)^\ell \,_{-2}Y_{\ell, -m}^* \right]^* \tilde{h}^*_{\ell, m}(-f)\ \\
    & = \frac{1}{2} \sum_{\substack{\ell\geq 2 \\ \ell \geq m > 0}} \,\left[ \,_{-2}Y_{\ell, m}^* + (-1)^\ell \,_{-2}Y_{\ell, -m} \right] \tilde{h}^*_{\ell, m}(-f) \ ,
    \end{aligned}
\end{equation}
where, in the last line, we have used that only $m>0$ modes have frequency content at positive frequencies, which allows us to only evaluate the $m>0$ modes in both the SPA and FFT part of the waveform. One analogously finds for the cross polarization
\begin{equation}
    \begin{aligned}
    \tilde{h}_{\times}(f) =-\frac{i}{2} \sum_{\substack{\ell\geq 2 \\ \ell \geq m > 0}} \, \Big[ & \,_{-2}Y_{\ell, m} + (-1)^{\ell+1} \, _{-2}Y_{\ell, -m} \Big] \\[-2.5ex] & \times \tilde{h}^*_{\ell, m}(-f)\ ,
    \end{aligned}
\end{equation}
by following the same steps starting from Eq.~\eqref{eq:hcross_td}.

\bibliography{Bibliography}

\end{document}